\documentclass[preprint,review,10pt]{elsarticle}

\usepackage[utf8]{inputenc}
\usepackage{amsmath,amssymb}
\usepackage{graphicx}
\usepackage{booktabs}
\usepackage{listings}
\usepackage{xcolor}
\usepackage{tikz}
\usetikzlibrary{positioning,arrows.meta,fit,backgrounds,calc}
\usepackage{hyperref}

\journal{ArXiV}

\begin{document}

\begin{frontmatter}

\title{A C++ implementation of the G-Scheme stiff ODE solver
with multi-resolution sparse hash-table kernel lookup}

\author[sap]{Riccardo Malpica Galassi}
\author[sap]{Mauro Valorani\corref{cor1}}
\cortext[cor1]{Corresponding author.}
\address[sap]{Mechanical and Aerospace Engineering Department, \\
Sapienza University of Rome, Rome, Italy}

\begin{abstract}
The G-Scheme is an explicit, adaptive numerical integration framework for
stiff systems of ordinary differential equations that exploits a local
time-scale decomposition provided by the eigensystem of the Jacobian of the
vector field (the \emph{kernel set}). Its computational cost is dominated by
the evaluation of a fresh kernel set at each time step. In a recent paper
[Malpica Galassi \& Valorani, Combust.\ Theory Model.\ 2025] we introduced a
multi-resolution sparse hash-table strategy that replaces the online kernel
computation with retrieval from a precomputed lookup table, demonstrated
with a Python implementation. Here we present \textsc{cpp-gscheme}, a
production-grade C++ implementation of both the G-Scheme and the
multi-resolution hash table, and document the porting methodology that
guarantees behavioural equivalence with the Python reference: identical
scaling, binning and hashing semantics (verified key-by-key), a
bitwise-inert retrieval hook in the integrator, and a zero-Python evaluation
path for the chemical source terms. On 33 mechanisms for n-heptane
autoignition -- a skeletal family of 56--459 species plus the detailed
654-species mechanism -- the hash-tabulated C++ G-Scheme retrieves 100\% of
the kernel sets at every mechanism size, performing zero online
eigendecompositions, and outruns the natively-compiled CVODE reference by a
factor that grows with the mechanism size: from 2.7$\times$ at 56 species
to 9.3$\times$ on the detailed mechanism, with observed cost scaling of
$\sim N^{1.4}$ against CVODE's $\sim N^{1.9}$. A three-metric accuracy
assessment (ignition delay, equilibrium state, and an accumulated error
evaluated in entropy progress) shows that the solver contributes
negligibly to the error budget, which is set by the skeletal reduction
alone; on the resulting error--cost Pareto, a 5\% ignition-delay budget is
met by the 80-species skeletal integrated 225$\times$ faster than CVODE on
the detailed mechanism.
The software, training-set builders and benchmark drivers are released as
open source.
\end{abstract}

\begin{keyword}
stiff ODE integration \sep computational singular perturbation \sep
G-Scheme \sep hash table \sep tabulation \sep combustion kinetics \sep C++
\end{keyword}

\end{frontmatter}

\section*{Program summary}
\begin{small}
\noindent
\emph{Program Title:} cpp-gscheme (hash-table release)\\
\emph{Developer's repository link:} \url{https://github.com/mauvalora/cpp-gscheme-ode}\\
\emph{Licensing provisions:} MIT\\
\emph{Programming language:} C++14/17 (core, adapters), Python 3 (drivers,
training-set construction), pybind11 bindings\\
\emph{External routines/libraries:} Eigen 3, pybind11, Cantera 3
(chemical kinetics), NumPy, Matplotlib\\[2pt]
\emph{Nature of problem:} Numerical integration of stiff ODE systems arising
from finite-rate chemical kinetics. Explicit CSP-based solvers such as the
G-Scheme remove stiffness by projecting out fast exhausted modes, but
require the eigensystem of the Jacobian (the kernel set) at every step, a
cost scaling as $N^2$--$N^3$ with the number of unknowns.\\[2pt]
\emph{Solution method:} The kernel set is retrieved from a multi-resolution
sparse hash table populated offline (or dynamically) from training
trajectories. Selected state variables are Box--Cox scaled, binned at
multiple resolutions $\epsilon = 2^{-n_{\rm exp}}$, and hashed with a
polynomial hash function; retrieval probes resolutions finest-to-coarsest
and accepts a stored kernel when the scaled-state distance is below a
threshold. On a hit, both the Jacobian evaluation and the
eigendecomposition are skipped; only the mode amplitudes are recomputed.
The C++ implementation adds a zero-Python evaluation path for the source
terms and is verified to be behaviourally identical to the published Python
reference.\\[2pt]
\emph{Additional comments:} Includes training-set builders, equivalence and
integration test suites, and benchmark drivers reproducing the published
validation campaign.
\end{small}

\tableofcontents
\section{Introduction}
\label{sec:intro}

Stiff systems of ordinary differential equations (ODEs) arising from
finite-rate chemical kinetics are conventionally integrated with implicit
solvers such as CVODE~\cite{cvode}, whose robustness comes at the price of
Jacobian factorisations within Newton iterations. The
G-Scheme~\cite{ValoraniPaolucci2009} is an explicit alternative rooted in
Computational Singular Perturbation (CSP)
theory~\cite{LamGoussis,GoussisLam}: at each step the tangent space is
decomposed -- via the eigensystem of the Jacobian of the vector field --
into fast (exhausted), active, and slow (dormant) subspaces, and only the
active dynamics is integrated explicitly, with algebraic corrections
accounting for the fast and slow contributions. The decomposition data --
eigenvalues and right/left eigenvectors, collectively the \emph{kernel set}
-- must be refreshed along the trajectory, and its computation dominates
the cost of the method.

In the method paper~\cite{MalpicaValorani2025} we introduced a
multi-resolution sparse hash-table strategy that replaces the online kernel
computation with retrieval from a lookup table populated offline from
training trajectories (or dynamically during integration), and demonstrated
-- with a pure-Python implementation -- a ten- to twenty-fold speed-up over
a traditional CSP solver, together with robustness to the retrieval of
approximate kernel sets \cite{Carinci2026}. 

The Python implementation, however, left the
central practical question open: can the G-Scheme equipped with the hash
table outperform a state-of-the-art, natively compiled implicit solver in
absolute terms?

This paper answers that question affirmatively by presenting
\textsc{cpp-gscheme}, a C++ implementation of the complete stack -- the
G-Scheme integrator, the multi-resolution hash table, the offline
training-set builder, and model adapters coupling directly to
libCantera~\cite{cantera} -- and documenting two aspects of independent
interest to developers of scientific software:

\begin{enumerate}
\item a \emph{porting methodology} that pins the C++ implementation to the
Python reference at the level of individual numerical semantics (rounding
modes, modular arithmetic, container behaviour), verified by key-level and
trajectory-level equivalence tests (Section~\ref{sec:porting});
\item a \emph{benchmark campaign} on 33 mechanisms for n-heptane
autoignition -- a skeletal family of 56--459 species plus the detailed
654-species mechanism -- in which the hash-tabulated C++ G-Scheme achieves
a 100\% kernel retrieval rate at every size, eliminating the
eigendecomposition from the time loop entirely, and overtakes CVODE by a
margin that grows with the problem dimension, up to 9.3$\times$ on the
detailed mechanism (Section~\ref{sec:performance}).
\end{enumerate}

The remainder of the paper is organised as follows.
Sections~\ref{sec:gscheme} and~\ref{sec:hashtable} summarise the G-Scheme
and the multi-resolution hash-table strategy, referring the reader
to~\cite{ValoraniPaolucci2009,ValoraniEtAl2018,MalpicaValorani2025} for the
complete formulations. Section~\ref{sec:software} describes the software
architecture. Section~\ref{sec:porting} presents the porting methodology
and its verification. Section~\ref{sec:performance} reports the benchmark
campaign. Section~\ref{sec:conclusions} concludes.

\section{The G-Scheme in brief}
\label{sec:gscheme}

The G-Scheme~\cite{ValoraniPaolucci2009,ValoraniEtAl2018} advances the state
$\mathbf{x}(t)\in\mathbb{R}^N$ of the stiff system
$\dot{\mathbf{x}} = \mathbf{g}(\mathbf{x})$ by decomposing, at each step,
the tangent space into a fast (\emph{tail}), an \emph{active}, and a slow
(\emph{head}) subspace spanned by right eigenvectors $\mathbf{a}_i$ of the
Jacobian $\mathbf{J} = \partial\mathbf{g}/\partial\mathbf{x}$, ordered by
decreasing eigenvalue magnitude, with dual (left) eigenvectors
$\mathbf{b}^i$ satisfying
$\langle\mathbf{a}_i,\mathbf{b}^j\rangle=\delta_i^j$. The tail dimension
$T$ collects exhausted fast modes whose amplitudes
$f^i = \mathbf{b}^i\!\cdot\!\mathbf{g}$ satisfy a user tolerance; the head
dimension $H$ collects dormant slow modes. Only the $H-T$ active modes are
integrated -- here with an explicit fourth-order Runge--Kutta scheme
projected onto the active subspace -- with a time step of the order of the
fastest \emph{active} scale, typically orders of magnitude larger than the
fastest scale of the system. Quasi-linear algebraic corrections account for
the tail (and optionally head) dynamics. The information required at each
step -- $(\lambda_i, \mathbf{a}_i, \mathbf{b}^i)$, the \emph{kernel set} --
is what the hash table provides.

\section{The multi-resolution sparse hash table in brief}
\label{sec:hashtable}

The lookup strategy of~\cite{MalpicaValorani2025} maps a user-selected
subset (mask) of $k \ll N$ state variables -- typically the temperature and
a handful of major species -- to a stored kernel set:

\begin{enumerate}
\item \emph{Scaling}: each masked variable is Box--Cox transformed,
$y_i = (|x_i|^{\lambda}-1)/\lambda$ with $\lambda = 0.3$, then min--max
normalised to the unit interval using bounds recorded from the training
set;
\item \emph{Binning}: the scaled state is discretised at multiple
resolution levels, $\mathbf{g} =
\mathrm{round}(\mathbf{y}_{\rm scaled}/\epsilon_{n_{\rm exp}}) + 1$ with
$\epsilon_{n_{\rm exp}} = 2^{-n_{\rm exp}}$,
$n_{\rm exp} \in \{3,\dots,10\}$;
\item \emph{Hashing}: a polynomial hash
$\mathrm{key} = \sum_i g_i\, l^{\,i} \bmod s$ (base $l=131$, prime modulus
$s$) produces the key into a two-layer table
$\mathrm{table}[n_{\rm exp}][\mathrm{key}] \Rightarrow$ kernel set.
\end{enumerate}

Retrieval probes the levels finest-to-coarsest and accepts a stored kernel
when $\lVert\mathbf{y}_{\rm scaled} -
\mathbf{y}_{\rm scaled}^{\rm stored}\rVert \le \tau$ (the published
campaign used $\tau = 0.1$). Coarse levels aggregate many states per bin
and act as robust fall-backs when the queried state is far from the
training data; the G-Scheme tolerates the resulting approximate kernels by
automatically reducing its time step, without loss of accuracy. On a miss
the kernel is computed on the fly and, optionally, inserted into the table
(\emph{dynamic expansion}).

\section{Software architecture}
\label{sec:software}

\textsc{cpp-gscheme} is organised in four layers
(Fig.~\ref{fig:architecture}):

\begin{figure}[htbp]
\centering
\resizebox{\textwidth}{!}{%
\begin{tikzpicture}[
  font=\small,
  box/.style={draw, rounded corners=2pt, align=center, fill=#1,
              inner sep=5pt, minimum height=8mm},
  hot/.style={-{Stealth[length=2.6mm]}, very thick, red!75!black},
  setup/.style={-{Stealth[length=2.4mm]}, thick, dashed, black!60},
  lab/.style={font=\scriptsize, align=center, fill=white, inner sep=1.5pt},
]
\node[box=blue!8, minimum width=32mm] (builder)
  {training-set builder\\ \scriptsize CVODE sampling +\\
   \scriptsize \texttt{compute\_kernel}};
\node[box=blue!8, right=8mm of builder, minimum width=30mm] (drivers)
  {benchmark drivers\\ \scriptsize campaigns, accuracy,\\ \scriptsize plots};
\node[box=blue!8, right=8mm of drivers, minimum width=26mm] (loader)
  {table loader\\ \scriptsize mask, scaling,\\ \scriptsize bulk insert};
\begin{scope}[on background layer]
\node[draw, dashed, rounded corners, fill=blue!3, inner sep=3mm,
      fit=(builder)(drivers)(loader),
      label={[anchor=north west, font=\scriptsize\bfseries]north west:
             Python drivers (orchestration only)}] (pylayer) {};
\end{scope}

\node[box=orange!12, below=22mm of builder.south west, anchor=north west,
      minimum width=60mm, xshift=-2mm] (core)
  {\textbf{\texttt{\_core}: StandaloneGSchemeC}\\[1pt]
   \scriptsize retrieval hook $\to$ (hit: adopt $\lambda,\mathbf{A},\mathbf{B}$
   $\mid$ miss: FD $\mathbf{J}$ + eig)\\
   \scriptsize $\to$ findM/findH $\to$ projected RK4 $\to$ corrections};
\node[box=orange!12, right=14mm of core, minimum width=38mm] (table)
  {\textbf{MultiResolution}\\\textbf{HashTable}\\[1pt]
   \scriptsize $\mathrm{table}[n_{\rm exp}][\mathrm{key}] \Rightarrow$
   kernel set\\
   \scriptsize Box--Cox $\to$ bin $\to$ hash};
\begin{scope}[on background layer]
\node[draw, rounded corners, fill=orange!4, inner sep=3.5mm,
      fit=(core)(table),
      label={[anchor=south east, font=\scriptsize\bfseries]south east:
             C++ core}] (corelayer) {};
\end{scope}
\draw[hot, <->] (core) -- node[lab]{retrieve\\/ insert} (table);

\path (pylayer.south) -- (corelayer.north) coordinate[midway] (bmid);
\draw[black!45, dotted, thick]
  ([xshift=-4mm]bmid -| corelayer.west) --
  ([xshift=4mm]bmid -| corelayer.east);
\node[font=\scriptsize\itshape, fill=white, inner sep=1pt,
      anchor=west] at ($(bmid -| corelayer.west)+(-15mm,2.4mm)$)
  {pybind11 boundary};

\node[box=green!10, below=17mm of core.south, anchor=north,
      minimum width=44mm, xshift=10mm] (cantera)
  {\textbf{CanteraReactorAdapter}\\
   \scriptsize $[T, Y_1 \ldots Y_{N_s}]$, cached $p$};
\node[box=green!10, right=8mm of cantera, minimum width=28mm] (toy)
  {Van der Pol,\\ Chem3, \ldots};
\begin{scope}[on background layer]
\node[draw, rounded corners, fill=green!3, inner sep=3.5mm,
      fit=(cantera)(toy),
      label={[anchor=south east, font=\scriptsize\bfseries]south east:
             \texttt{gscheme\_adapters}: AdapterBase}] (adplayer) {};
\end{scope}

\node[box=gray!15, below=11mm of cantera.south, anchor=north,
      minimum width=44mm] (libct)
  {\textbf{libCantera} (native C++)\\
   \scriptsize thermodynamics, kinetics};

\draw[hot] ([xshift=10mm]core.south) -- node[lab, pos=0.5, left=2.5mm]
  {\textcolor{red!75!black}{\texttt{rhs(y,t)} -- NativeRhs capsule}\\
   \textcolor{red!75!black}{no GIL, no numpy, no Python frame}}
  (cantera.north);
\draw[hot] (cantera) -- node[lab, right=1mm]{native calls} (libct);

\draw[setup] (drivers.south) -- node[lab, pos=0.55, left=2mm]
  {\texttt{set\_integrator, attach\_hash\_table,}\\
   \texttt{set\_native\_rhs, integrate\_to()}}
  (drivers.south |- corelayer.north);
\draw[setup] (loader.south) to[out=-75, in=75]
  node[lab, pos=0.7, right=1.5mm]{offline training\\ (bulk insert)}
  (table.north);
\draw[setup] (builder.west) to[out=180, in=180, looseness=1.4]
  node[lab, pos=0.35, left=1mm]{construct,\\ \texttt{set\_TPX}}
  (adplayer.west);
\end{tikzpicture}%
}
\caption{The four-layer architecture of \textsc{cpp-gscheme}. Python
orchestrates (dashed arrows: one-time setup and data exchange across the
pybind11 boundary); the per-step hot path (solid red) -- integrator,
kernel retrieval, and source-term evaluation down to libCantera -- is a
single native call stack, free of the interpreter.}
\label{fig:architecture}
\end{figure}
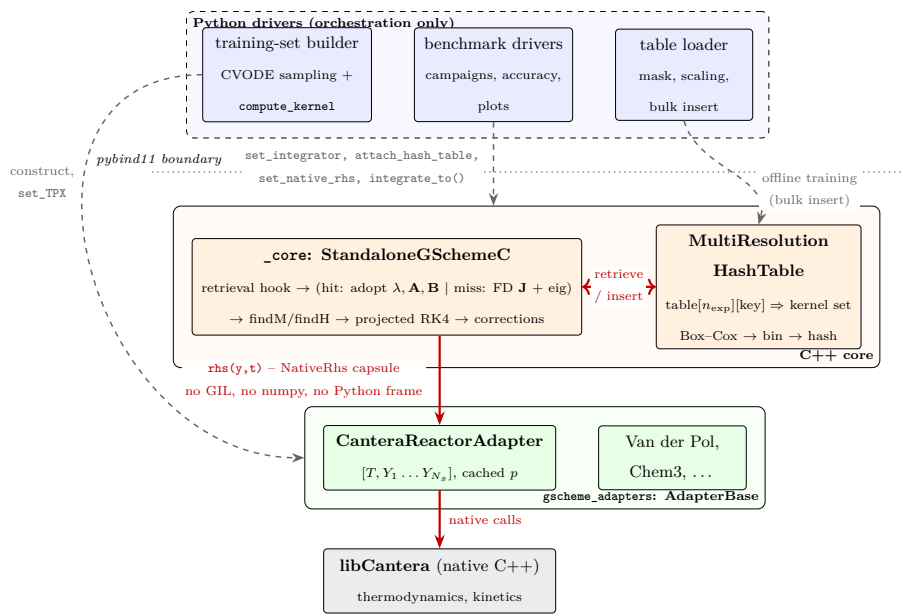

\paragraph{C++ core (\texttt{cpp\_src/})} The integrator
\texttt{StandaloneGSchemeC} exposes a \texttt{scipy}-like interface
(\texttt{set\_initial\_value}, \texttt{integrate},
\texttt{integrate\_to}) through pybind11. Each step performs: CSP kernel
acquisition (Section~\ref{sec:hook}); tail/head dimension selection;
projected RK4 integration of the active modes; tail (and optionally head)
corrections. Eigen~3 provides the dense linear algebra; the
eigendecomposition follows the LAPACK-backed \texttt{Eigen::EigenSolver}
with eigenvalue ordering and complex-conjugate-pair realification matching
the Python reference.

\paragraph{Hash table (\texttt{cpp\_src/hash\_table.\{hpp,cpp\}})} The class
\texttt{MultiResolutionHashTable} implements the strategy of
Section~\ref{sec:hashtable} with
$\texttt{table}[n_{\rm exp}][\mathrm{key}] \rightarrow
\texttt{shared\_ptr<KernelEntry>}$; one \texttt{KernelEntry}
(scaled state, $\lambda$, $\mathbf{A}$, $\mathbf{B}$, stored SIM dimension)
is shared by all resolution levels, reproducing the reference-counting
semantics -- and the memory footprint -- of the Python dictionaries, which
store references to shared NumPy arrays. Retrieval performs one $O(1)$
probe per level; the measured cost is 1--2\,$\mu$s per lookup, below 1\% of
a step.

\paragraph{Model adapters (\texttt{cpp\_adapters/})} All physics is
delivered through a minimal abstract interface, detailed in
Section~\ref{sec:adapters}; the Cantera adapter couples the core directly
to the libCantera binaries, with no Python in the evaluation path.

\paragraph{Python drivers} Training-set construction
(\texttt{database\_builder\_cpp.py}: CVODE sampling of the training
trajectories; kernel sets computed by the core's own
\texttt{compute\_kernel} so that tabulated and online kernels are
numerically consistent), table loading, benchmark and plotting drivers.

\subsection{Model adapters: one interface for Cantera and non-Cantera
models}
\label{sec:adapters}

The integrator is model-agnostic: every physical model enters through the
abstract class \texttt{AdapterBase}, whose essential contract is four
methods over raw C arrays (no framework types cross the interface):

\begin{lstlisting}[language=C++]
struct AdapterBase {
  virtual void rhs(const double* y, double t, double* dydt) = 0;
  virtual int  n_variables() const = 0;
  virtual int  n_elements()  const { return 0; }  // conserved invariants
  virtual bool has_analytical_jacobian() const { return false; }
  // optional: jacobian(), get_initial_state(), variable_names(), ...
};
\end{lstlisting}

\texttt{n\_elements} declares the number of conserved linear invariants
(the atomic elements for reacting mixtures), which bounds the head-side
mode selection; \texttt{has\_analytical\_jacobian} lets a model supply an
exact Jacobian, with the core falling back to finite differences
otherwise. Two families of adapters are provided.

\paragraph{Non-Cantera models.} Analytic test problems (Van der Pol,
a three-species fast/slow kinetics model) implement the interface in a
few dozen lines each and serve three roles: unit-test fixtures for the
integrator (the hash-table integration tests of
Section~\ref{sec:porting} run on Van der Pol), stiffness benchmarks with
known closed-form spectra, and templates for user-defined models.

\paragraph{The Cantera reactor adapter.} \texttt{CanteraReactorAdapter}
implements the homogeneous constant-pressure (or constant-volume) reactor
with the state convention $\mathbf{y} = [T, Y_1, \dots, Y_{N_s}]$. Its
construction loads the mechanism once
(\texttt{Cantera::newSolution}) and caches the thermodynamic and kinetics
handles, the molecular weights, and -- crucially for the isobaric
formulation -- the reactor pressure, so that each \texttt{rhs} call is a
pure sequence of native library calls:
\texttt{setMassFractions\_NoNorm} and \texttt{setState\_TP} to set the
thermodynamic state, \texttt{getNetProductionRates} and the partial molar
enthalpies to assemble
$\dot Y_k = \dot\omega_k W_k / \rho$ and
$\dot T = -\sum_k h_k \dot\omega_k W_k / (\rho c_p)$.
The mass fractions are deliberately \emph{not} renormalised inside
\texttt{rhs}: the finite-difference Jacobian perturbs one species at a
time, and renormalisation would contaminate the perturbed column with
$O(\epsilon)$ changes to every other species. Mixture initialisation
supports both a direct composition string and the
fuel/oxidiser/equivalence-ratio specification, delegating the
stoichiometry to Cantera.

Each adapter is exposed to Python through numpy-based pybind11 bindings
(so the same object also serves SciPy's BDF reference driver), and to the
core through the zero-Python bridge of Section~\ref{sec:native} -- one
adapter instance, two coupling levels.

\subsection{Robustness on large mechanisms}
\label{sec:robustness}

Scaling the port from the small skeletal mechanisms to the 654-species
detailed mechanism exposed a failure mode specific to realistic ignition
problems: at a fresh mixture state most mass fractions are \emph{exactly}
zero (651 of 654 species for stoichiometric n-heptane/air), and a
finite-difference Jacobian built with a single scalar perturbation is
rank-deficient -- its eigendecomposition then yields clusters of
numerically zero eigenvalues with linearly dependent eigenvectors, and
the right-eigenvector matrix cannot be inverted for the dual basis.
Three complementary measures address this:

\begin{enumerate}
\item \emph{Per-component perturbations.} The Jacobian column for
variable $i$ uses $\epsilon_i = \max(\epsilon_{\rm rel}\,|y_i|,
\epsilon_{\rm abs})$ instead of a global scalar: temperature
($\sim$$10^3$~K) and trace species ($\sim$$10^{-20}$) are perturbed on
their own scales. Beyond enabling the large mechanisms, the improved
Jacobian quality shortens runs across the family (e.g.\ the 103-species
case drops from 421 to 296 steps, as larger, better-founded time steps
are selected).
\item \emph{Initial-condition floor.} An optional floor
(\texttt{--y-floor}, typically $10^{-20}$) replaces exactly-zero species
in the \emph{initial condition only}, making the very first Jacobian
well-conditioned. The placement matters: an earlier variant that clamped
species inside the finite-difference loop altered the Jacobian at every
step and degraded a 296-step case to 4688 steps before crashing --
whereas the one-time floor at a physically negligible value is inert
(no-floor and floored runs coincide on mechanisms that do not need it).
\item \emph{Graceful singular-basis fallback.} When the right-eigenvector
matrix is nevertheless singular (detected by rank-revealing LU,
\texttt{FullPivLU::isInvertible}), the kernel computation reports failure
instead of resorting to an $O(N^3)$ SVD pseudo-inverse; the integrator
then reuses the previous step's basis and refreshes only the mode
amplitudes $\mathbf{f} = \mathbf{B}\,\mathbf{g}$ -- legitimate because
the kernel varies on the slow scale, the same property that underlies the
hash table itself. The offline training-set builder applies the analogous
policy: states with singular eigendecompositions are simply not
tabulated.
\end{enumerate}

A fourth, minor safeguard: the adaptive step growth (capped at $1.5\times$
per step) is additionally bounded by an absolute \texttt{dt\_max},
preventing runaway steps when nearly all modes are exhausted near
equilibrium.

\subsection{The retrieval hook}
\label{sec:hook}

The hash table enters the integrator at the point where the kernel would
be refreshed. Crucially, when a table is attached the (finite-difference)
Jacobian evaluation is \emph{deferred} until the lookup is known to miss:
a hit skips both the $N{+}1$ evaluations of the source term needed to form
$\mathbf{J}$ and the $O(N^3)$ eigendecomposition, and recomputes only the
mode amplitudes $\mathbf{f} = \mathbf{B}\,\mathbf{g}(\mathbf{x})$ and the
timescales:

\begin{lstlisting}
if (need_basis_update) {
  entry = hash_table->retrieve(y);          // O(1) per level
  if (entry) {                              // hit:
    lambda = entry->lambda;                 //   adopt tabulated kernel
    A = entry->A;  B = entry->B;
    f = B * rhs;  tau = timescales(lambda); //   refresh amplitudes only
  } else {                                  // miss:
    J = finite_diff_jacobian(y);            //   deferred until here
    compute_csp_kernel(J, rhs, lambda, A, B, f);
    if (dynamic_insert) hash_table->insert(y, lambda, A, B, M);
  }
}
\end{lstlisting}

By construction the miss path executes the identical arithmetic of the
table-free solver; we verify in Section~\ref{sec:porting} that attaching an
empty table leaves trajectories bitwise unchanged.
\clearpage
\subsection{The zero-Python evaluation path}
\label{sec:native}

The core and the adapters are separate compiled modules. In the baseline
coupling, each source-term evaluation crosses the boundary through a
one-line Python lambda, paying for the interpreter lock, two array
conversions and one interpreter frame per call ($\sim$5--15\,$\mu$s). The
native-RHS bridge eliminates this: only C types cross the module boundary,
\begin{lstlisting}[language=C++]
struct NativeRhs {
  void (*fn)(void* ctx, const double* y, double t, double* dydt);
  void* ctx;   // opaque adapter pointer
  int n_vars;
};
\end{lstlisting}
and every call site (step source term, finite-difference Jacobian columns,
RK4 stages, corrections, offline kernel construction) inherits the direct
path. With the bridge installed, the complete evaluation chain -- core
$\to$ adapter $\to$ libCantera -- is a single native call stack, and the
comparison with Cantera's CVODE (itself native C++ driven from Python) is
level. Trajectories are bitwise identical to the Python-callback path.
Measured end-to-end, the bridge contributes a further 6--13\% at 88
species -- the physics and linear algebra dominate at realistic sizes --
but it renders the hot loop GIL-free, a prerequisite for the multi-cell
(PDE) parallelisation planned as future work.

\section{Porting methodology and verification}
\label{sec:porting}

Porting a numerical method between language ecosystems is not a
translation exercise: NumPy/LAPACK and C++/Eigen differ in rounding
conventions, modular arithmetic, eigenvector sign and ordering
conventions, and container semantics, and each difference can silently
change the computed answer. The port therefore pins the C++ semantics to
the Python reference at the level of individual operations -- first for
the integrator itself, then for the hash table, whose value lies entirely
in its exact behaviour (a single bin or key computed differently changes
hit patterns and, silently, performance claims) -- and verifies the
equivalence mechanically.

\subsection{Solver-level semantics: mode selection and conjugate-pair
atomicity}
\label{sec:modeselection}

The most delicate part of the integrator port is the pair of
mode-selection routines. \texttt{findM} scans the modes from fastest to
slowest, accumulating into a running vector $\delta\mathbf{w}$ the state
change that projecting out each mode would leave unresolved, and returns
the first index at which any component exceeds its error weight
$\mathrm{ewt}_k = \mathrm{rtol}\,|y_k| + \mathrm{atol}$; \texttt{findH}
performs the mirrored backward scan for the dormant slow modes. Two
implementation subtleties are decisive:

\paragraph{Exact cumulative semantics.} The scan must reproduce the
reference implementation's cumulative sums \emph{exactly}: an early C++
variant that accumulated complex-conjugate pairs atomically inside the
forward scan produced cumulative values in a different order than the
reference (which follows a plain \texttt{cumsum}), and hence different
tail dimensions, different step sizes, and diverging trajectories on some
mechanisms. The final \texttt{findM} accumulates strictly one mode at a
time and applies the pair constraint \emph{retroactively}: if the
threshold is first exceeded at index $j$ and modes $j{-}1, j$ form a
conjugate pair, the boundary is backed up to $j-1$ so the pair stays
intact on the active side.

\paragraph{Sign-convention invariance.} Complex conjugate pairs
$\lambda = \sigma \pm i\omega$ must never straddle a subspace boundary --
splitting a pair yields a non-physical real basis and a diverging
integration. The backward scan of \texttt{findH} is additionally
sensitive to a subtler portability issue: the \emph{individual}
contribution of a pair member, $\tfrac12\,\Delta t^2\,
\mathbf{a}_j f^j |\lambda_j|$, depends on the sign convention of the
complex eigenvectors, and Eigen and LAPACK make different (both
legitimate) choices. \texttt{findH} therefore accumulates the two members
of a pair \emph{atomically} before testing the threshold: the net pair
contribution is invariant to the convention, and the C++ and Python
implementations select identical head dimensions. The asymmetry of the
two strategies -- retroactive adjustment forward, atomic accumulation
backward -- is thus not stylistic but forced: one preserves
cumulative-sum equality, the other sign-convention invariance. The same
pair bookkeeping protects the boundary cases (a pair occupying the last
two indices of either scan). These fixes turned mechanisms that
previously crashed or diverged (notably the 138-species member of the
family) into routine runs.

\subsection{Hash-table semantic pinning}

On the table side, three semantic traps deserve documentation, as they
would silently break key-level equivalence:

\begin{enumerate}
\item \emph{Rounding.} NumPy's \texttt{np.round} implements
round-half-to-even (banker's rounding); C's \texttt{round()} rounds half
away from zero. The binning step uses \texttt{std::rint} under the default
\texttt{FE\_TONEAREST} mode, which matches NumPy exactly.
\item \emph{Modular arithmetic.} Python's \texttt{\%} is a floored modulo
(non-negative results for positive moduli) whereas the C++ \texttt{\%}
truncates towards zero; out-of-range queried states produce negative bins,
so the polynomial hash emulates the Python semantics explicitly.
\item \emph{Payload sharing.} The Python table stores, at every resolution
level, \emph{references} to the same NumPy arrays. The C++ table stores
\texttt{shared\_ptr}s to a single \texttt{KernelEntry}; a per-level deep
copy would multiply the memory footprint by the number of levels (several
GB for the larger mechanisms).
\end{enumerate}

\subsection{Equivalence test suite}

Two test programs, shipped with the code, certify the port:

\paragraph{Table equivalence
(\texttt{tests/test\_hash\_table\_equivalence.py})} builds the Python
reference table and the C++ table from the same synthetic combustion-like
training set (2000 states, 6-variable mask, $n_{\rm exp}=6\dots10$) and
verifies: bitwise-identical scaled states; identical hash keys for all
10\,000 (state, level) pairs; identical per-level occupancy (including
collision-overwrite patterns); and identical hit/miss outcome, matched
resolution level, and payload for $\sim$8000 queries spanning training
states, midpoints, perturbed and far (miss) states.

\paragraph{Integrator equivalence
(\texttt{tests/test\_hash\_gscheme\_integration.py})} verifies on a Van der
Pol problem that (i) attaching an \emph{empty} table leaves the solver
trajectory \emph{bitwise identical} to the table-free solver -- the
retrieval hook is inert on misses; (ii) dynamic insertion populates the
table during integration while preserving accuracy; (iii) an
in-distribution table yields a 100\% hit rate with zero kernel
computations.

\subsection{Replication of the published campaign}

As an end-to-end check, the C++ stack replicates the method paper's
validation on the 88-species skeletal mechanism with the published
settings: the classic C++ G-Scheme takes 364 steps against the published
362; the hash-tabulated run retrieves 357/357 kernel sets In-Distribution
(published: all states retrieved) and 481/481 Out-Of-Distribution
(published: 491 steps, all retrieved), with the retrieval resolution
levels shifting coarser Out-Of-Distribution exactly as published.

\section{Performance}
\label{sec:performance}

\subsection{Campaign definition}

Constant-pressure autoignition of stoichiometric n-heptane/air
($T_0 = 1000$~K, $p = 1$~atm, $t_{\rm end} = 0.1$~s), over 33 mechanisms: a
family of 32 skeletal mechanisms (56--459 species) derived from the
654-species detailed mechanism~\cite{Mehl2011} with the simplification
procedure of~\cite{MalpicaSimplification}, plus the detailed mechanism
itself. For the two largest mechanisms the training trajectories were
subsampled (every 2nd state for $N_s = 459$, every 4th for $N_s = 654$) to
respect the 18~GB memory of the benchmark machine; the retrieval rate was
unaffected. Solvers: CVODE (Cantera's native
\texttt{ReactorNet}); the classic C++ G-Scheme (kernel computed every
step); the hash-tabulated C++ G-Scheme with an In-Distribution table (one
training trajectory at $T_0$) and an Out-Of-Distribution table (two
bracketing trajectories at $T_0 \pm 15$~K, the harder scenario of the
method paper). Settings as published: mask
\{nC$_7$H$_{16}$, O$_2$, HCO, H$_2$O, CO$_2$, $T$\}, $n_{\rm exp}=3\dots10$,
acceptance tolerance $0.1$; G-Scheme tolerances
rtol$_{\rm tail}=10^{-3}$, atol$_{\rm tail}=10^{-9}$,
rtol$_{\rm head}=10^{-4}$, atol$_{\rm head}=10^{-10}$, $\gamma = 0.2$.
Timings are process CPU times of the integration loop only, on an Apple M3
Pro (18~GB, Apple clang 21, Cantera 3.0.1); table construction and training
are excluded, as in the method paper.
\subsection{Results}

\begin{figure}[htbp]
\centering
\includegraphics[width=0.8\textwidth]{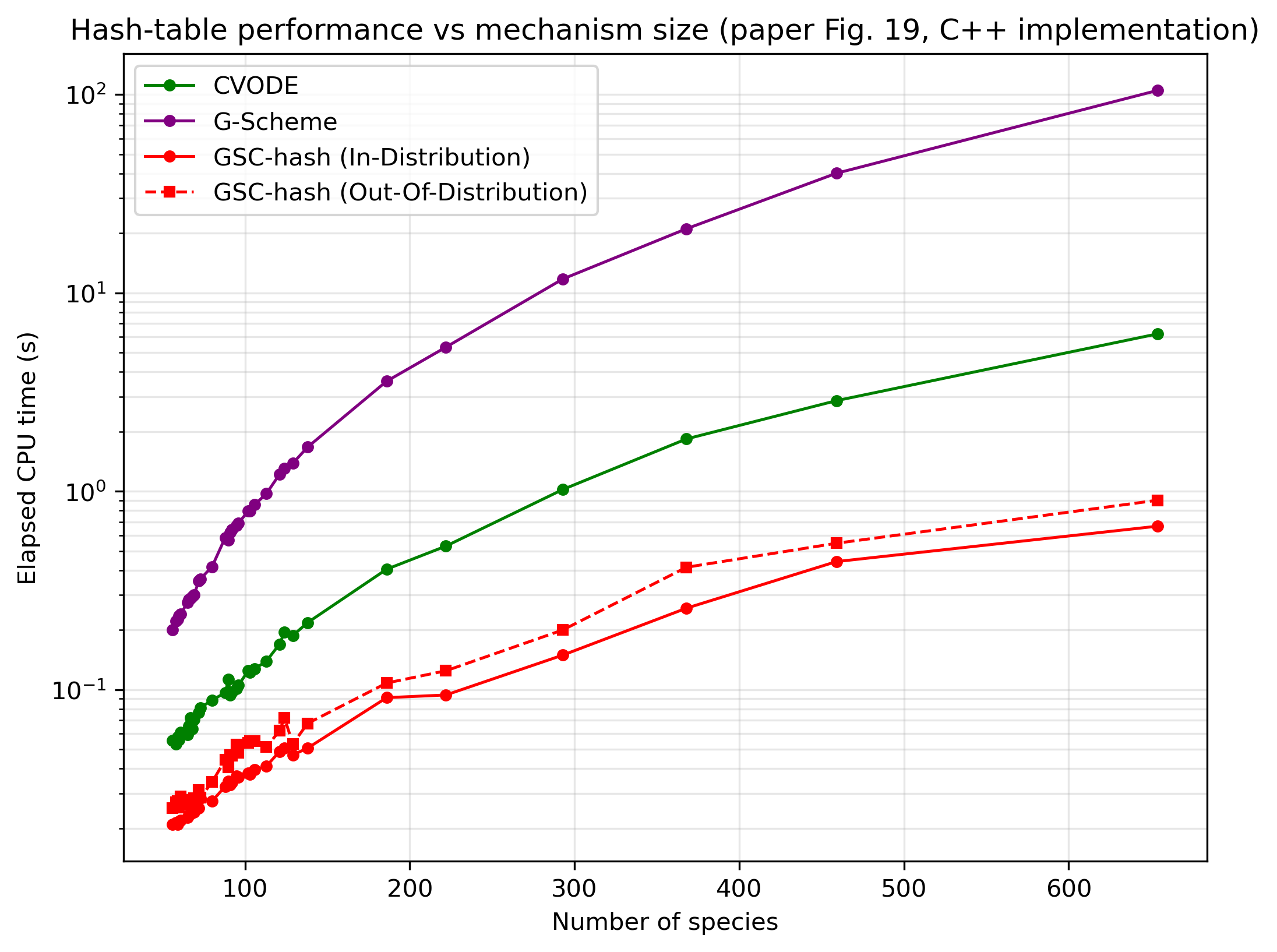}
\caption{Elapsed CPU time against the number of species: CVODE (green),
classic C++ G-Scheme (purple), hash-tabulated C++ G-Scheme with
In-Distribution (red, solid) and Out-Of-Distribution (red, dashed) tables.}
\label{fig:fig19cpp}
\end{figure}

\begin{figure}[htbp]
\centering
\includegraphics[width=0.8\textwidth]{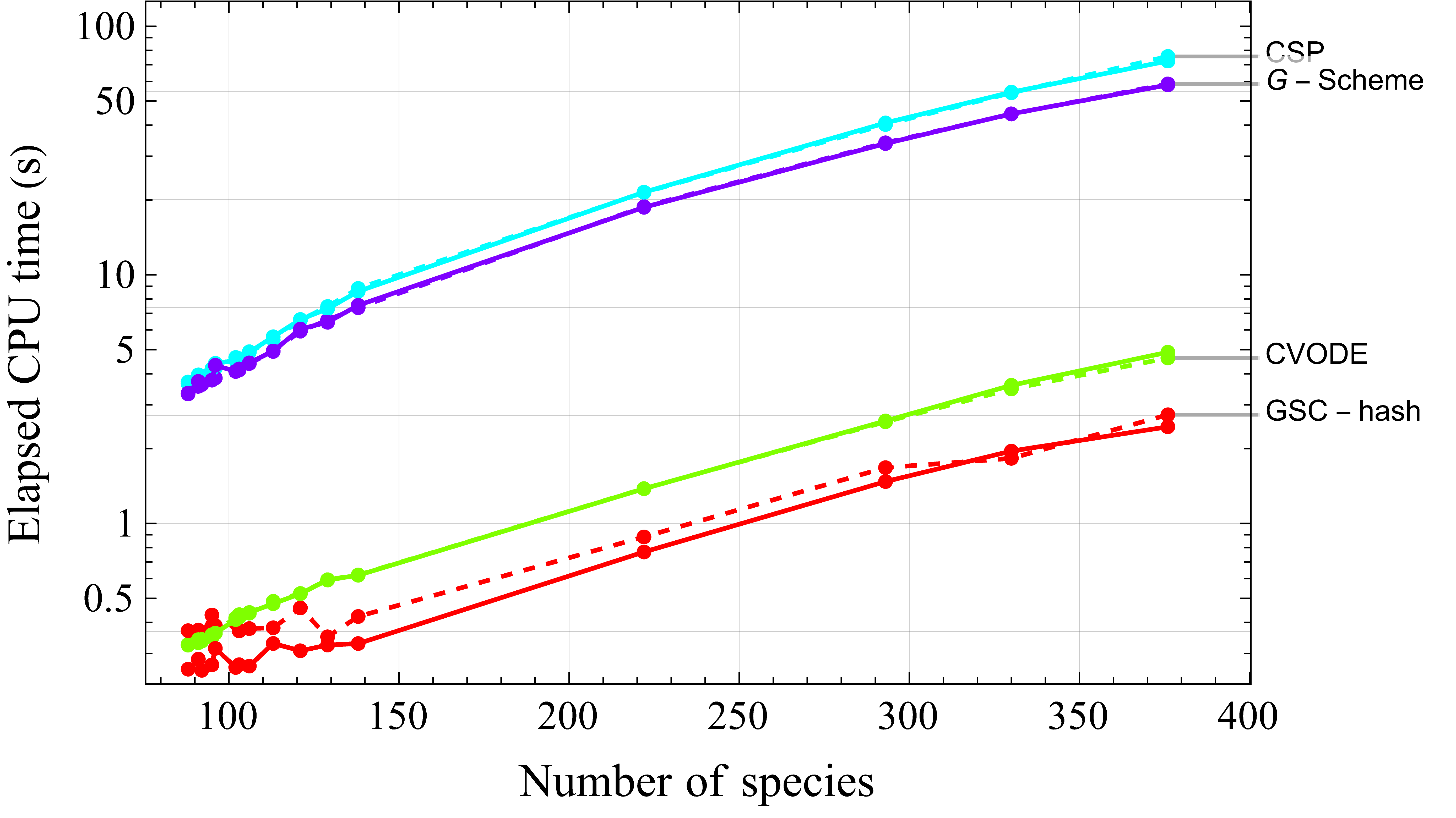}
\caption{For comparison: the corresponding campaign of the method
paper~\cite{MalpicaValorani2025} with the Python implementation (CSP solver
cyan, G-Scheme purple, CVODE green, hash-tabulated G-Scheme red).}
\label{fig:fig19python}
\end{figure}

Across all 33 mechanisms and both training scenarios the retrieval rate is
100\%: \emph{the hash-tabulated solver performs zero online
eigendecompositions at every mechanism size}, up to and including the
detailed 654-species mechanism. Table~\ref{tab:campaign} and
Fig.~\ref{fig:fig19cpp} collect the timings;
Fig.~\ref{fig:fig19python} reproduces the corresponding figure of the
method paper's Python implementation for comparison.

\begin{table}[htbp]
\centering
\caption{Elapsed CPU time (integration loop) for CVODE, the classic C++
G-Scheme, and the hash-tabulated C++ G-Scheme with In-Distribution (ID) and
Out-Of-Distribution (OOD) tables. All hash runs: 100\% retrieval, zero
online kernel evaluations.}
\label{tab:campaign}
\tiny
\begin{tabular}{rrrrrrrr}
\toprule
$N_s$ & $N_r$ & CVODE [s] & G-Scheme [s] & hash ID [s] & hash OOD [s] &
$\frac{\rm CVODE}{\rm ID}$ & $\frac{\hbox{\scriptsize G-Scheme}}{\rm ID}$\\
\midrule
56 & 543 & 0.062 & 0.20 & 0.021 & 0.025 & 3.0 & 9.9 \\
58 & 558 & 0.059 & 0.23 & 0.022 & 0.028 & 2.7 & 10.5 \\
59 & 560 & 0.065 & 0.23 & 0.022 & 0.027 & 3.0 & 10.7 \\
60 & 562 & 0.062 & 0.24 & 0.022 & 0.026 & 2.8 & 10.9 \\
61 & 564 & 0.067 & 0.25 & 0.022 & 0.029 & 3.1 & 11.3 \\
65 & 592 & 0.066 & 0.28 & 0.023 & 0.029 & 2.9 & 12.4 \\
66 & 626 & 0.073 & 0.29 & 0.024 & 0.028 & 3.0 & 11.7 \\
67 & 634 & 0.079 & 0.30 & 0.024 & 0.027 & 3.2 & 12.3 \\
68 & 640 & 0.069 & 0.30 & 0.026 & 0.028 & 2.7 & 11.7 \\
69 & 645 & 0.075 & 0.30 & 0.024 & 0.028 & 3.0 & 12.2 \\
72 & 681 & 0.080 & 0.35 & 0.025 & 0.031 & 3.2 & 14.0 \\
73 & 700 & 0.086 & 0.35 & 0.028 & 0.029 & 3.0 & 12.4 \\
80 & 752 & 0.093 & 0.43 & 0.028 & 0.034 & 3.3 & 15.3 \\
88 & 705 & 0.101 & 0.58 & 0.032 & 0.046 & 3.1 & 18.1 \\
90 & 869 & 0.118 & 0.56 & 0.032 & 0.041 & 3.7 & 17.5 \\
91 & 748 & 0.101 & 0.63 & 0.034 & 0.049 & 3.0 & 18.5 \\
92 & 750 & 0.104 & 0.65 & 0.034 & 0.048 & 3.1 & 19.2 \\
95 & 780 & 0.109 & 0.69 & 0.038 & 0.053 & 2.9 & 18.5 \\
96 & 784 & 0.111 & 0.71 & 0.037 & 0.049 & 3.0 & 19.1 \\
102 & 867 & 0.134 & 0.81 & 0.038 & 0.057 & 3.5 & 21.1 \\
103 & 869 & 0.131 & 0.81 & 0.037 & 0.056 & 3.5 & 22.0 \\
106 & 887 & 0.138 & 0.88 & 0.040 & 0.057 & 3.4 & 21.9 \\
113 & 923 & 0.148 & 1.01 & 0.042 & 0.052 & 3.6 & 24.2 \\
121 & 996 & 0.176 & 1.25 & 0.047 & 0.066 & 3.7 & 26.3 \\
124 & 1251 & 0.206 & 1.34 & 0.053 & 0.079 & 3.9 & 25.3 \\
129 & 1065 & 0.207 & 1.45 & 0.053 & 0.057 & 3.9 & 27.2 \\
138 & 1140 & 0.224 & 1.69 & 0.053 & 0.070 & 4.2 & 31.9 \\
186 & 1877 & 0.410 & 3.68 & 0.095 & 0.115 & 4.3 & 38.8 \\
222 & 2088 & 0.548 & 5.33 & 0.091 & 0.129 & 6.0 & 58.6 \\
293 & 2645 & 1.031 & 11.71 & 0.150 & 0.203 & 6.9 & 78.1 \\
368 & 3209 & 1.870 & 21.68 & 0.266 & 0.456 & 7.0 & 81.5 \\
459 & 3755 & 2.925 & 40.05 & 0.372 & 0.591 & 7.9 & 107.7 \\
654 & 5258 & 6.339 & 105.80 & 0.599 & 0.839 & 10.6 & 176.8 \\
\bottomrule
\end{tabular}
\end{table}

Three observations:

\begin{enumerate}
\item \emph{The hash-tabulated C++ G-Scheme beats natively compiled CVODE
by a margin that grows with the problem size}: 2.7$\times$ at 56 species,
7.1$\times$ at 368 species, 9.3$\times$ on the detailed 654-species
mechanism (0.67~s against CVODE's 6.2~s), where it also outruns the
classic G-Scheme by 157$\times$. The observed cost scalings over the
56--654 range are $\sim N^{1.4}$ (hash), $\sim N^{1.9}$ (CVODE) and
$\sim N^{2.5}$ (classic G-Scheme).
\item \emph{Machine- and implementation-independent cross-validation}: the
classic-G-Scheme-to-CVODE elapsed-time ratio at $\sim$370 species is
11.5$\times$ in the published Python campaign and 11.5$\times$ here -- the
port preserves the algorithm's intrinsic cost structure exactly. By the
same normalisation the hash solver improves from 0.58$\times$ CVODE's time
(Python) to 0.14$\times$ (C++).
\item \emph{Out-Of-Distribution robustness}: with tables built only from
the $T_0 \pm 15$~K bracketing trajectories, retrieval remains 100\%; the
solver compensates for coarser retrieved kernels with more, smaller steps
(elapsed-time penalty 1.2--1.6$\times$), remaining 4.4$\times$ faster than
CVODE at 368 species.
\end{enumerate}

\subsection{How the speed-up arises: anatomy of a run}
\label{sec:anatomy}

\begin{figure}[htbp]
\centering
\begin{minipage}[t]{0.49\textwidth}
  \centering
  \includegraphics[width=\textwidth]{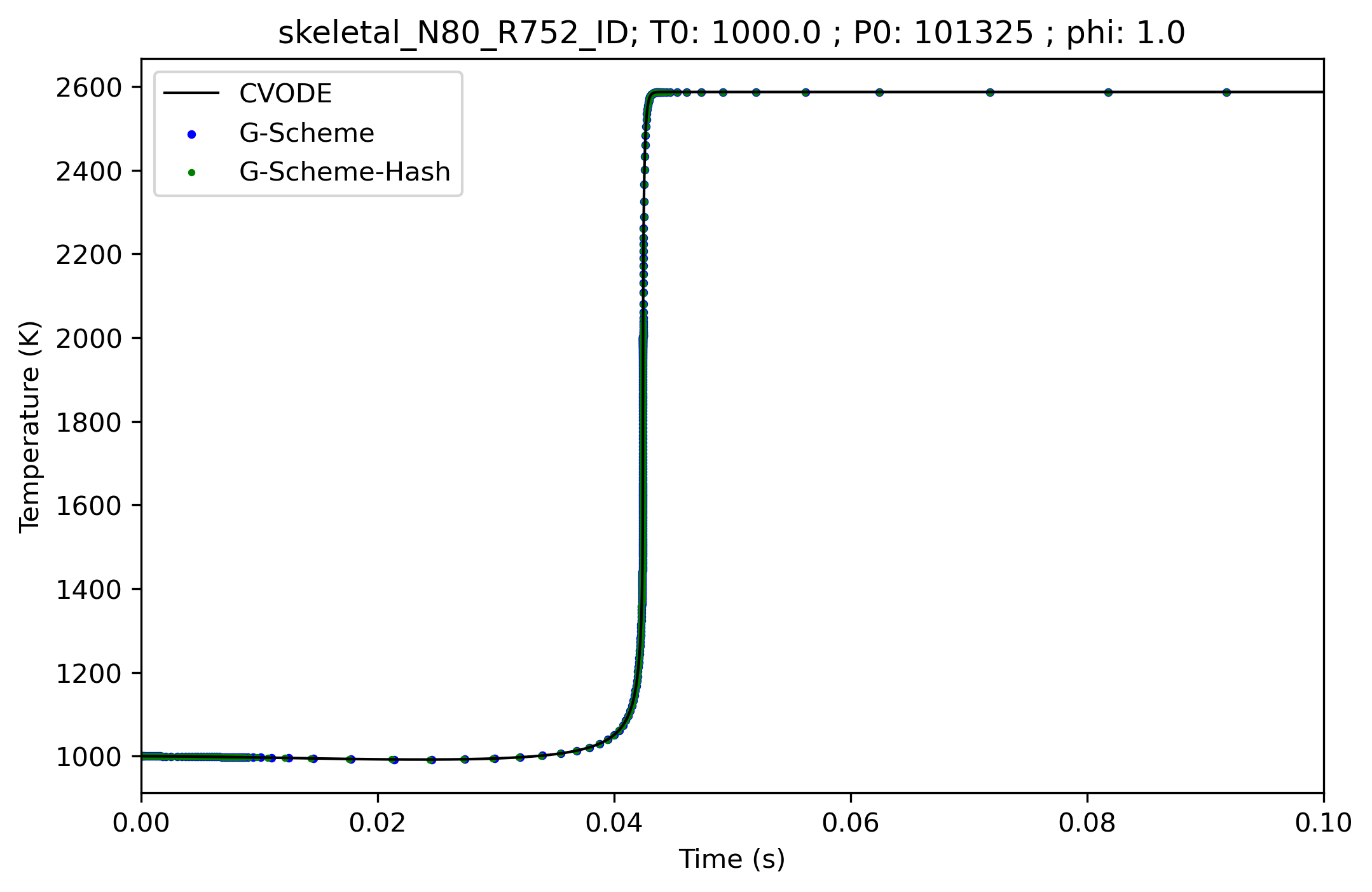}\\[-1pt]
  {\small (a)}
\end{minipage}\hfill
\begin{minipage}[t]{0.49\textwidth}
  \centering
  \includegraphics[width=\textwidth]{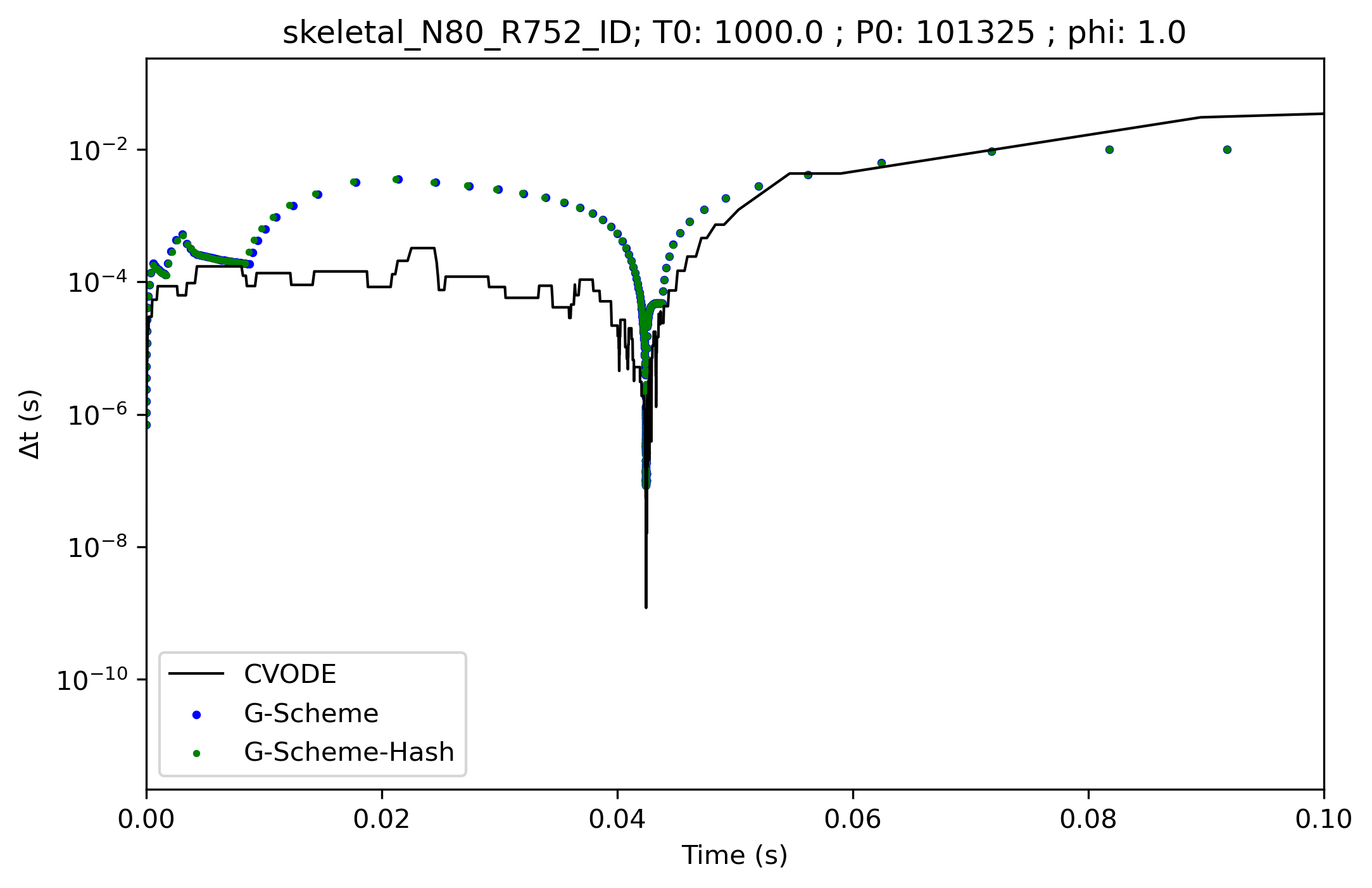}\\[-1pt]
  {\small (b)}
\end{minipage}\\[4pt]
\begin{minipage}[t]{0.49\textwidth}
  \centering
  \includegraphics[width=\textwidth]{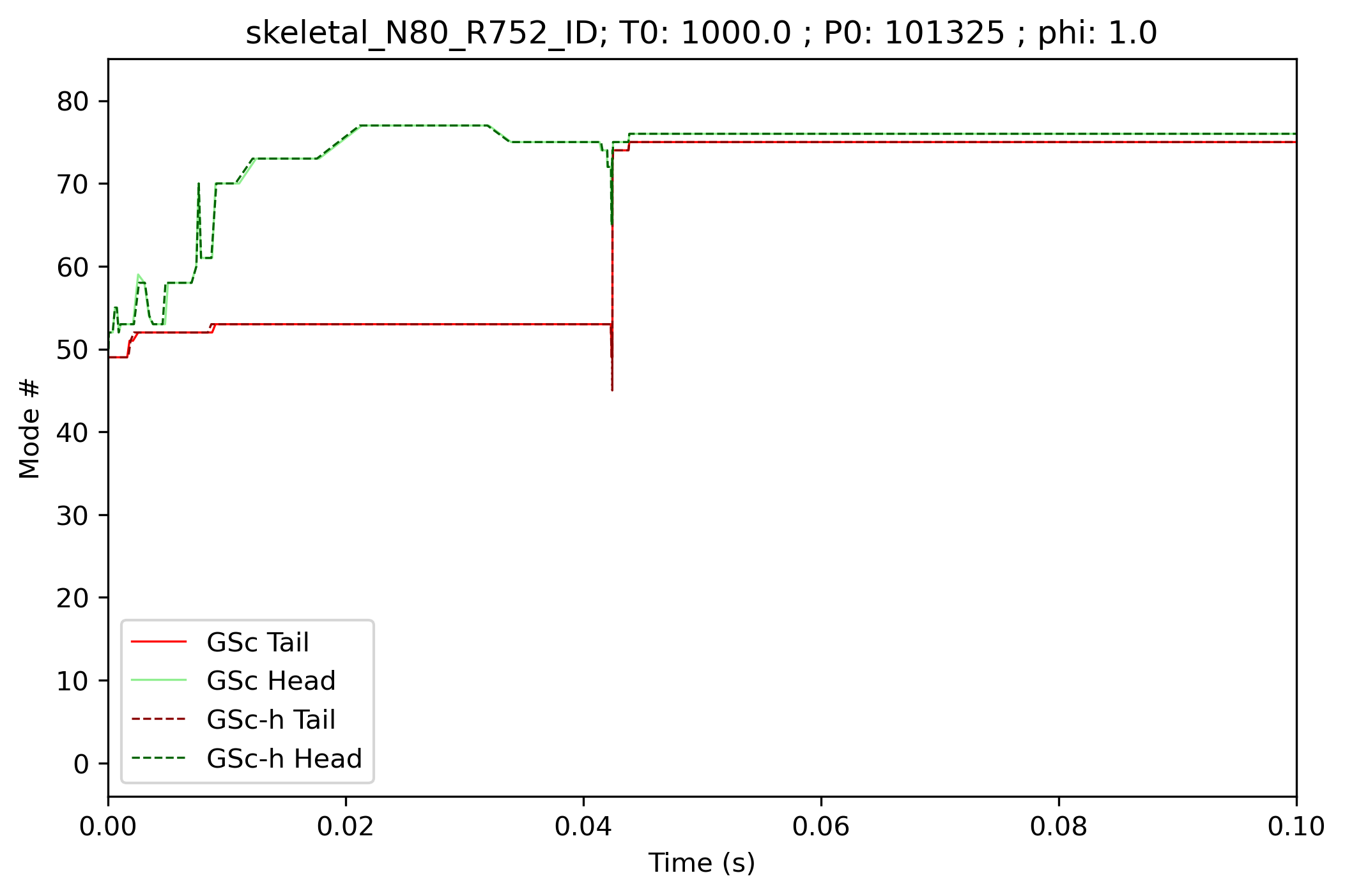}\\[-1pt]
  {\small (c)}
\end{minipage}\hfill
\begin{minipage}[t]{0.49\textwidth}
  \centering
  \includegraphics[width=\textwidth]{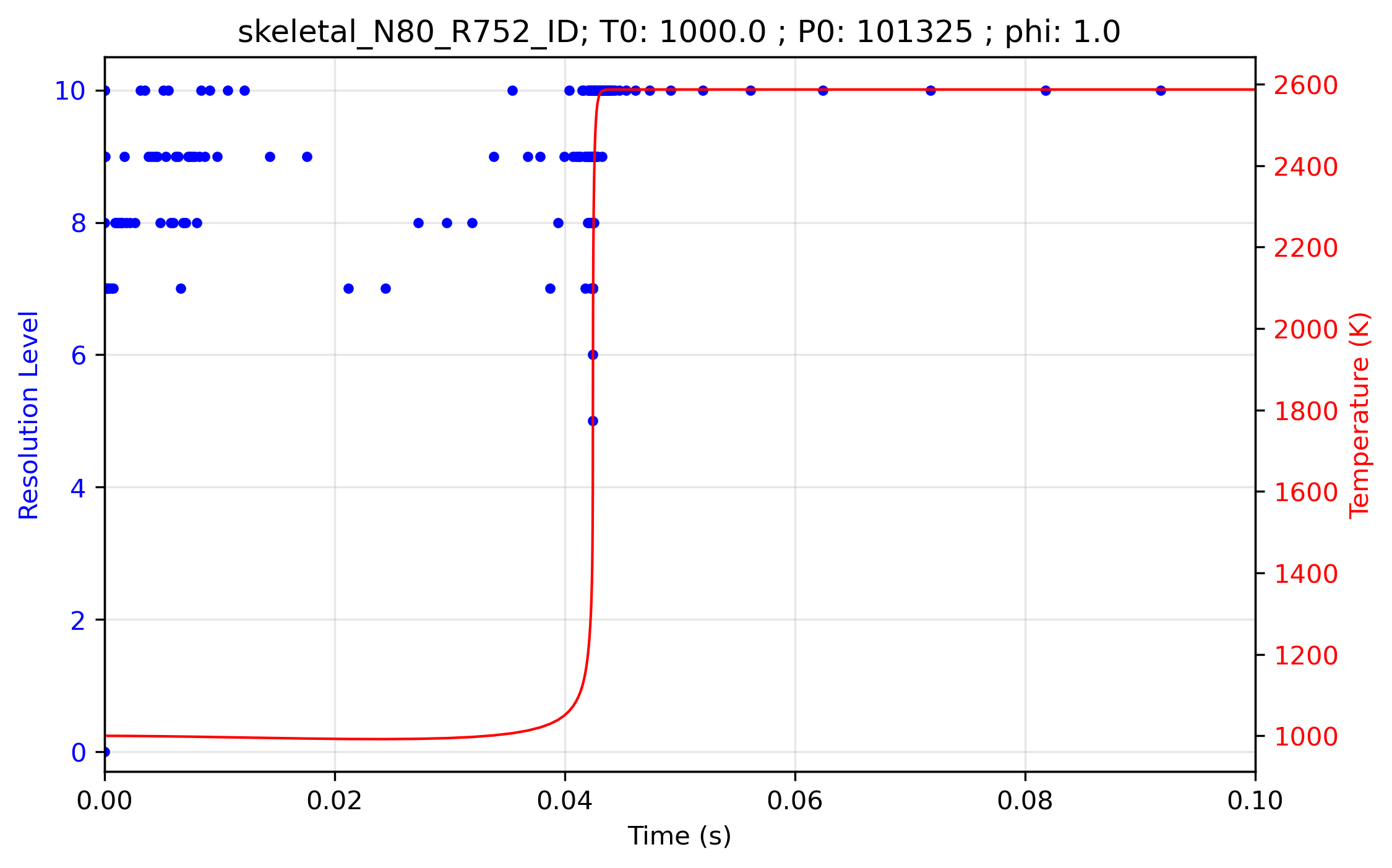}\\[-1pt]
  {\small (d)}
\end{minipage}
\caption{Anatomy of the 80-species In-Distribution run.
(a)~Temperature: CVODE, classic G-Scheme and hash-tabulated G-Scheme are
indistinguishable. (b)~Step size: the G-Scheme advances with steps one to
two orders of magnitude larger than CVODE throughout the induction period;
both refine at the ignition front, and the hash-tabulated steps (green)
exactly overlay the classic ones (blue). (c)~Tail and head dimensions: of
the $N=81$ unknowns, $\sim$50--53 fast modes are exhausted during
induction and $\sim$5--25 modes are active; after ignition the active
window narrows to 1--2 modes. The hash-tabulated dimensions (dashed)
overlay the classic ones. (d)~Resolution level of every kernel retrieval
(blue dots, left axis) with the temperature trace (red, right axis): all
316 steps retrieve, none computes.}
\label{fig:anatomy}
\end{figure}

There is no magic in the numbers of Table~\ref{tab:campaign}: the speed-up
is the product of three mechanisms, each of which can be observed directly
in the diagnostics of a single run. We use the 80-species skeletal
mechanism -- the optimum of the error--cost analysis of
Section~\ref{sec:accuracy} -- as the specimen
(Fig.~\ref{fig:anatomy}).

\paragraph{Mechanism 1: larger steps.} Panel~(b) contrasts the step-size
histories. CVODE, integrating the full stiff system, holds
$\Delta t \sim 10^{-4}$~s through the induction period; the G-Scheme,
having projected out the exhausted fast modes, selects the fastest
\emph{active} time scale and cruises at $\Delta t$ up to
$4\times10^{-3}$~s. Both solvers refine at the ignition front -- the
physics demands it -- and relax towards equilibrium. Net effect: 316
G-Scheme steps against 2399 CVODE steps (7.6$\times$ fewer).

\paragraph{Mechanism 2: fewer equations.} Panel~(c) shows the tail/head
dynamics behind those steps. During induction $\sim$50--53 of the 81 modes
are exhausted (tail) and the head sits at $\sim$73--77: the projected RK4
integrates only 5--25 active degrees of freedom -- an adaptive
model reduction performed on the fly, at every step, with the remaining
modes accounted for by the algebraic tail/head corrections. Past
ignition the active window collapses to 1--2 modes, and the solver rides
the slow manifold with near-maximal steps.

\paragraph{Mechanism 3: no kernel computation.} The price the classic
G-Scheme pays for panels (b) and (c) is the kernel set -- Jacobian plus
eigendecomposition at each of its 310 steps -- which is why, despite
taking 7.6$\times$ fewer steps, it is \emph{slower} than CVODE here
(0.432~s vs 0.093~s): the kernel consumes the entire step-count advantage.
Panel~(d) shows the hash table removing that price: every one of the 316
steps retrieves its kernel (at resolution levels 5--10, cost
$\sim$1~$\mu$s each), none computes, and the run completes in 0.028~s.
The overlay of the hash-tabulated curves on the classic ones in panels
(a)--(c) is the visual counterpart of the accuracy result of
Section~\ref{sec:accuracy}: retrieval changes neither the step selection
nor the subspace decomposition in any visible way.

\begin{figure}[htbp]
\centering
\includegraphics[width=0.62\textwidth]{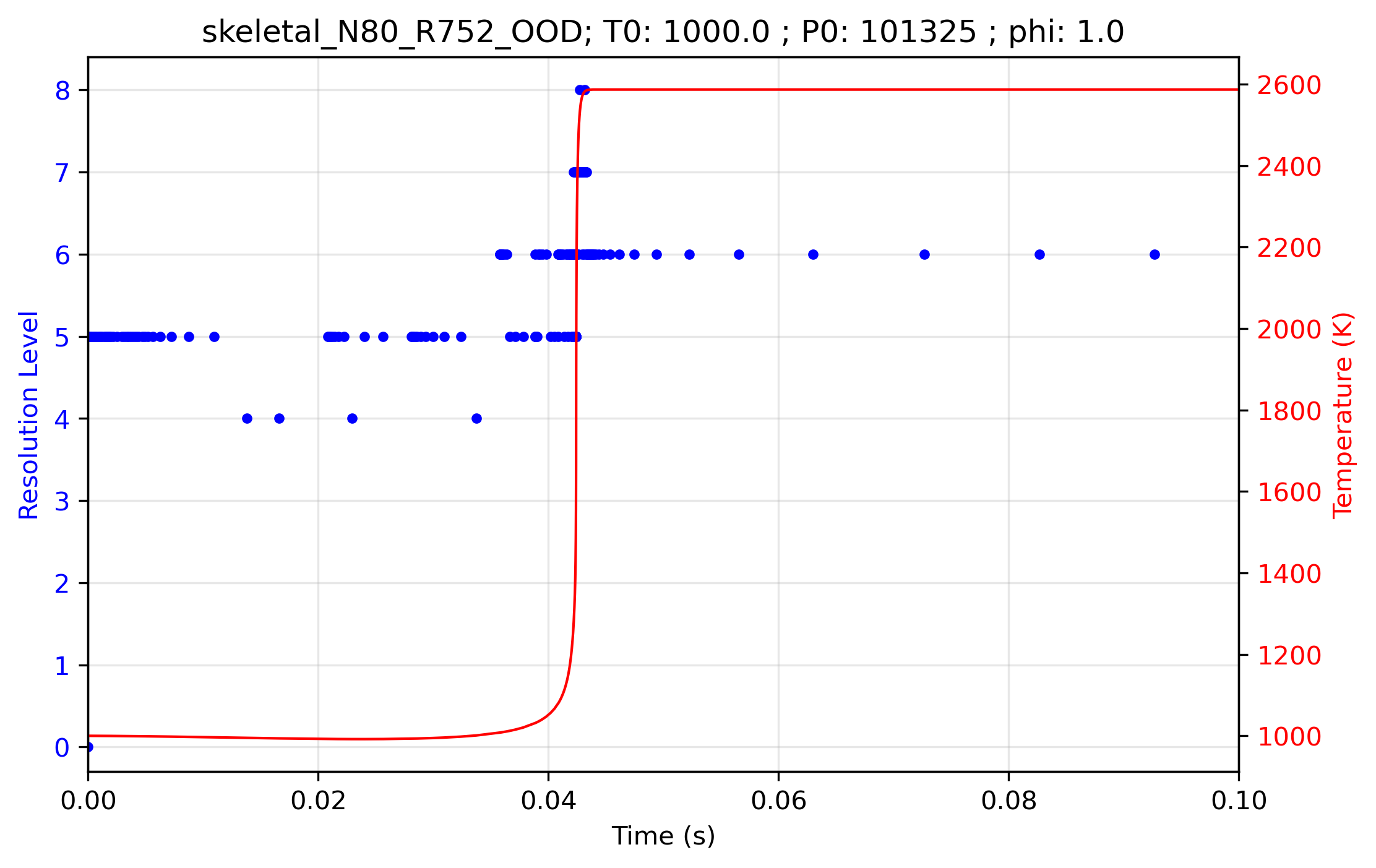}
\caption{Retrieval resolution levels of the Out-Of-Distribution run
(table trained only at $T_0 \pm 15$~K): the multi-resolution search
falls back to coarser levels (4--8 against the In-Distribution 5--10 of
Fig.~\ref{fig:anatomy}d), retrieval remains 100\% (376/376), and the
solver compensates with slightly smaller steps.}
\label{fig:anatomy_ood}
\end{figure}

Figure~\ref{fig:anatomy_ood} repeats panel~(d) for the
Out-Of-Distribution table: the retrieval shifts to coarser resolution
levels -- the multi-resolution structure absorbing the distance from the
training data -- while the hit rate stays at 100\% and the accuracy, per
Section~\ref{sec:accuracy}, is preserved through the step-size
self-protection.

\paragraph{Composing the speed-up.} The three mechanisms multiply, and
they compose with the offline skeletal reduction. For the 80-species
specimen: CVODE on the detailed mechanism (6.21~s) $\to$ CVODE on the
skeletal (0.093~s, a 67$\times$ reduction bought offline by CSP/TSR
simplification) $\to$ hash-tabulated G-Scheme on the skeletal (0.028~s, a
further 3.3$\times$ bought online by mechanisms 1--3). The product,
222$\times$, is the 225$\times$ of the Pareto analysis below to within
run-to-run timing noise -- every factor is accounted for.

\subsection{Accuracy}
\label{sec:accuracy}

Speed claims require certified accuracy. Three metrics are evaluated, all
as $L_2$ errors against a CVODE baseline: (i) the relative ignition-delay
error $e_\tau$, with $\tau_{\rm ign} = \arg\max_t \dot T$ located on a
monotone interpolant of the temperature history; (ii) equilibrium errors
(relative temperature error and $L_2$ mass-fraction error at
$t_{\rm end}$); and (iii) an accumulated trajectory error $E_{\rm traj}$
evaluated in normalised \emph{entropy progress}
$\xi = (s - s_0)/(s_{\rm eq} - s_0)$ -- the mixture entropy of an
adiabatic isobaric reactor increases monotonically, whereas the
temperature need not (e.g.\ under NTC chemistry), making $\xi$ the proper
independent variable for trajectory comparison. The metrics are applied
on two layers: a \emph{solver} layer, where each G-Scheme variant is
compared with the CVODE solution of the \emph{same} mechanism, and a
\emph{mechanism} layer, where each skeletal mechanism's CVODE solution is
compared with that of the detailed 654-species mechanism on the
intersection of their species sets (whose mass coverage of the detailed
equilibrium state is 1.0000 for every member of the family). The
\emph{combined} error of the reduction-plus-solver pipeline --
hash-tabulated G-Scheme on the skeletal mechanism against CVODE on the
detailed mechanism -- is measured directly as well.

\begin{figure}[htbp]
\centering
\includegraphics[width=0.85\textwidth]{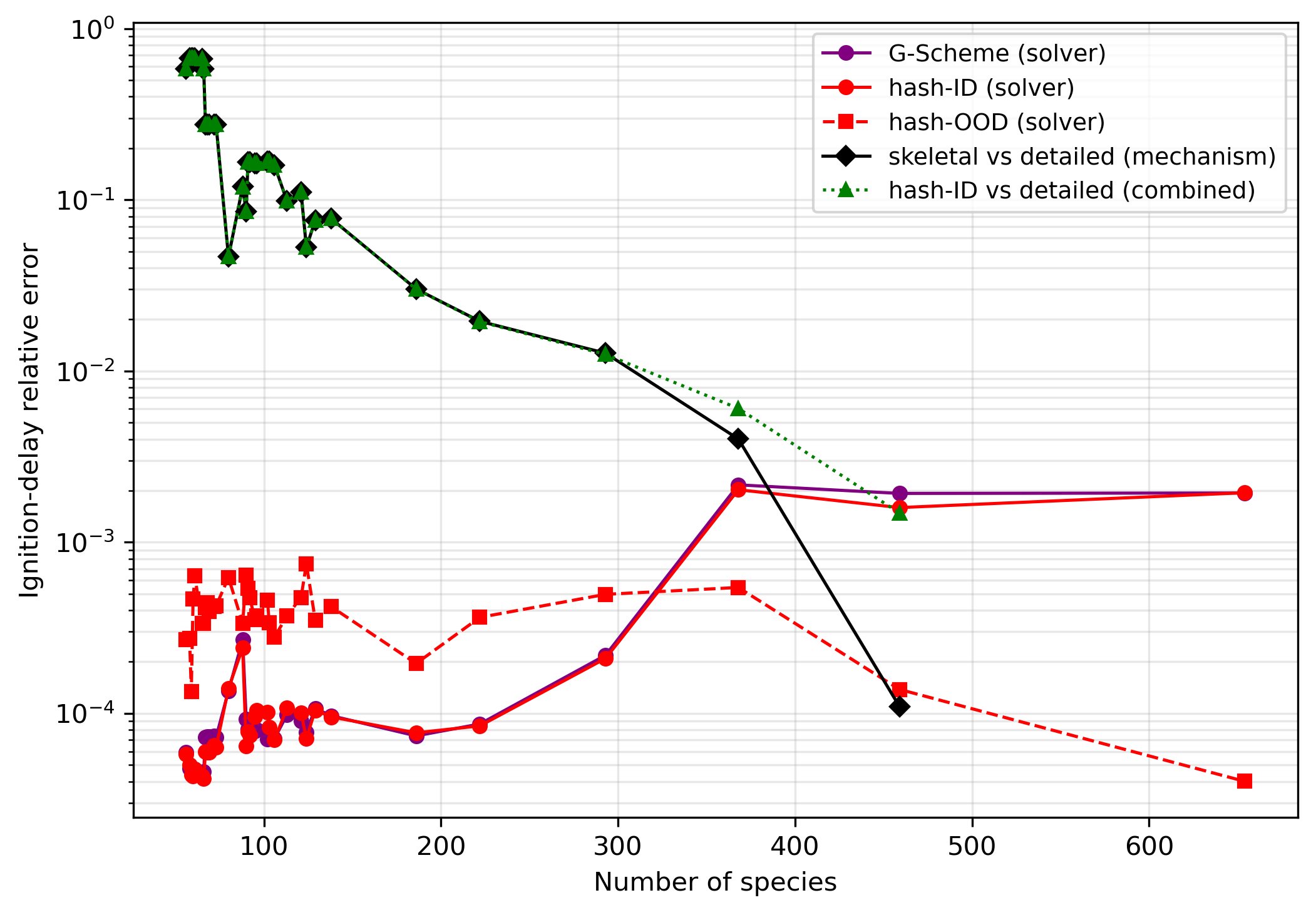}
\caption{Ignition-delay relative error against mechanism size, for the
solver layer (G-Scheme, hash-ID, hash-OOD vs same-mechanism CVODE), the
mechanism layer (skeletal vs detailed, black), and the combined pipeline
(hash-ID vs detailed, green dotted).}
\label{fig:accuracy_layers}
\end{figure}

Figure~\ref{fig:accuracy_layers} summarises the ignition-delay results
(the equilibrium and accumulated-error metrics behave analogously). Three
conclusions emerge:

\begin{enumerate}
\item \emph{The solver contributes negligibly to the error budget.} Below
300 species the solver-layer error is $e_\tau \lesssim 7\times10^{-4}$
(typically $10^{-4}$), two to three orders of magnitude below the
mechanism-reduction error; the combined error therefore coincides with
the mechanism error to three significant digits. A solver-error floor of
$e_\tau \approx 2\times10^{-3}$, common to the classic and hash-tabulated
variants and attributable to the integration tolerances, becomes visible
only beyond $\sim$400 species, where the reduction error drops below it.
\item \emph{Kernel retrieval costs no accuracy.} The hash-tabulated runs
match the classic G-Scheme's error at every size and for every metric --
in both training scenarios. The Out-Of-Distribution runs are, at the
largest sizes, \emph{more} accurate: the smaller steps triggered by
coarser retrieved kernels act as an implicit error control, the
robustness mechanism of the method paper manifesting as accuracy.
\item \emph{The error budget belongs entirely to the skeletal reduction},
which decays from $e_\tau \approx 0.6$ at 56 species to $10^{-4}$ at 459
(non-monotonically: the 80-species member is a favourable outlier at
$4.7\times10^{-2}$). The 88-species value (11.9\%) is consistent with the
corresponding assessment in the method paper, cross-validating the
present pipeline.
\end{enumerate}

\begin{figure}[htbp]
\centering
\includegraphics[width=0.85\textwidth]{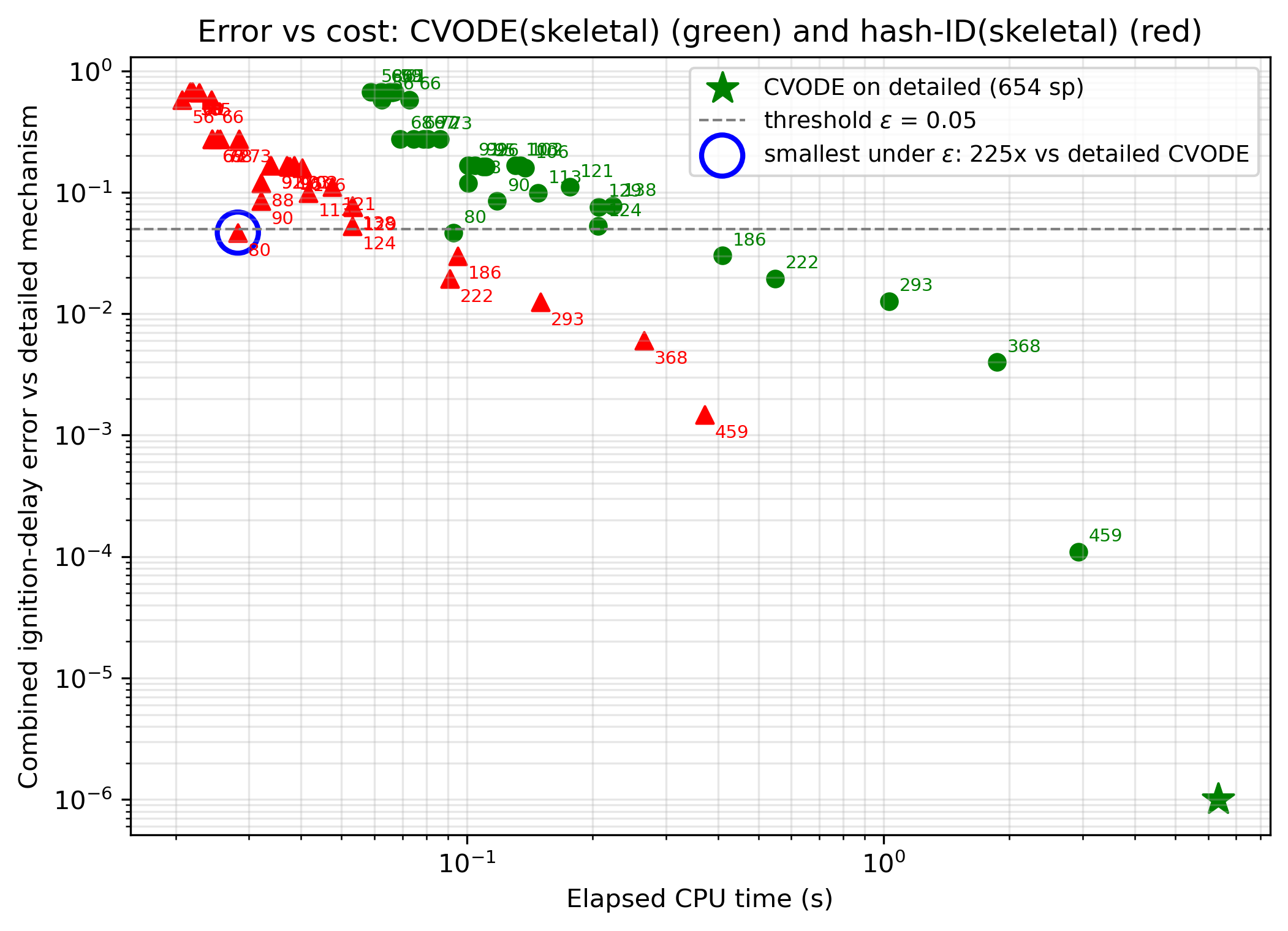}
\caption{Error--cost Pareto: combined ignition-delay error against
elapsed CPU time for CVODE on each skeletal mechanism (green circles) and
the hash-tabulated G-Scheme on the same mechanism (red triangles); green
star: CVODE on the detailed mechanism (reference, zero error). Dashed
line: user threshold $\varepsilon = 5\%$; the circled point is the
cheapest solution under threshold.}
\label{fig:pareto}
\end{figure}

Figure~\ref{fig:pareto} condenses the study into a single design chart:
at equal combined error the hash-tabulated points sit uniformly
5--9$\times$ to the left of the CVODE points, and the two reduction
avenues -- discarding species offline (moving down-left along either
curve) and discarding time scales online (jumping from green to red) --
compose multiplicatively. For an ignition-delay error budget of
$\varepsilon = 5\%$ the cheapest solution is the 80-species skeletal
integrated by the hash-tabulated G-Scheme in 0.028~s: \textbf{225$\times$
faster than CVODE on the detailed mechanism}. Tightening the budget moves
the optimum rightward along the red front ($\varepsilon = 2\%$: 222
species, 68$\times$; $\varepsilon = 1\%$: 368 species, 23$\times$) --
the practitioner chooses the accuracy, the chart returns the speed-up.

\section{Conclusions}
\label{sec:conclusions}

We presented \textsc{cpp-gscheme}, a C++ implementation of the G-Scheme
equipped with the multi-resolution sparse hash-table kernel lookup
of~\cite{MalpicaValorani2025}, together with the porting methodology --
semantic pinning of rounding, modular arithmetic and payload sharing,
enforced by key-level and trajectory-level equivalence tests -- that makes
the C++ table provably interchangeable with the Python reference. The
benchmark campaign establishes the result the method paper anticipated: with
kernel retrieval replacing online eigendecomposition, an \emph{explicit}
CSP-based solver outruns a natively compiled implicit BDF solver on stiff
ignition chemistry, by a factor that grows with the mechanism size. The
GIL-free integration loop opens the way to parallel multi-cell
(reaction--diffusion) applications, where all cells share a single table
and the trajectory-revisiting nature of turbulent reacting flows promises
high retrieval rates.


\section*{Data availability}
The source code, test suites, training-set builders, benchmark drivers and
the campaign records underlying the tables and figures are available in the
cpp-gscheme repository.

\bibliographystyle{elsarticle-num}

\appendix

\section{The tangential stretching rate as a training-set diagnostic}
\label{app:tsr}

The kernel sets stored in the hash table carry more information than the
integrator consumes: they are complete local time-scale decompositions of
the dynamics, and as such they support -- at no additional cost -- the
full CSP diagnostic toolbox. This appendix illustrates the point with the
\emph{tangential stretching rate} (TSR)~\cite{TSR2015}, and shows that,
beyond its diagnostic role, the TSR provides a practical quality check of
the training datasets themselves.

\paragraph{The TSR in brief.} The TSR is the amplitude-weighted stretching
rate that the dynamics exerts along its own trajectory,
\begin{equation}
\omega_\tau \;=\; \sum_i \hat w_i \,
  \mathrm{sign}\!\left(\mathrm{Re}\,\lambda_i\right) |\lambda_i|,
\qquad
\hat w_i \propto \left( \frac{|f^i|}{\lVert \mathbf{g} \rVert} \right)^{\!2},
\label{eq:tsr}
\end{equation}
with the weights of the exhausted modes ($i < M$) and of the conserved
modes set to zero, and the two members of a complex conjugate pair
carrying the pair modulus $\sqrt{(f^i)^2 + (f^{i-1})^2}$ -- the same
sign-convention-invariance device discussed in
Section~\ref{sec:modeselection}, here ensuring that the weights do not
depend on the eigenvector conventions of the linear-algebra backend. A
large positive $\omega_\tau$ identifies explosive (ignition-driving)
dynamics; a negative value, relaxation towards equilibrium. The TSR has
proven an incisive feature-tracking and cause/effect diagnostic in
laminar and turbulent reacting flows~\cite{TSR2015}; in the present
context, its computation from a training dataset requires only the
tabulated $(\lambda_i, \mathbf{b}^i, M)$ and one source-term evaluation
per stored state -- no integration.

\begin{figure}[htbp]
\centering
\includegraphics[width=0.8\textwidth]{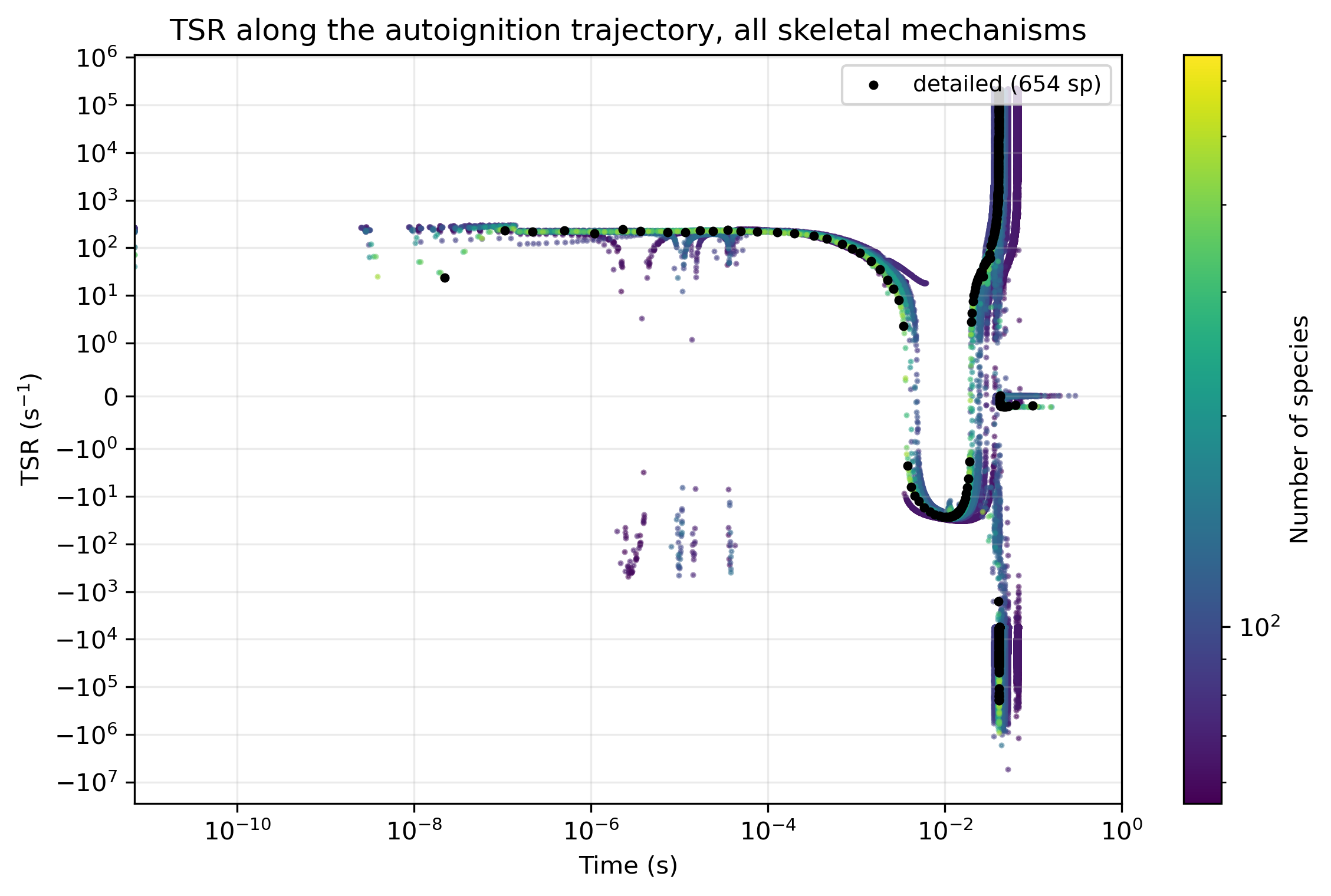}
\caption{TSR along the autoignition trajectory against logarithmic time,
for all 33 mechanisms of the family (colour: number of species; black:
detailed 654-species mechanism): the induction-period view. Note the
tightly bundled plateau over eight decades of time and the common
pre-ignition negative basin.}
\label{fig:tsr_time}
\end{figure}

\begin{figure}[htbp]
\centering
\includegraphics[width=0.8\textwidth]{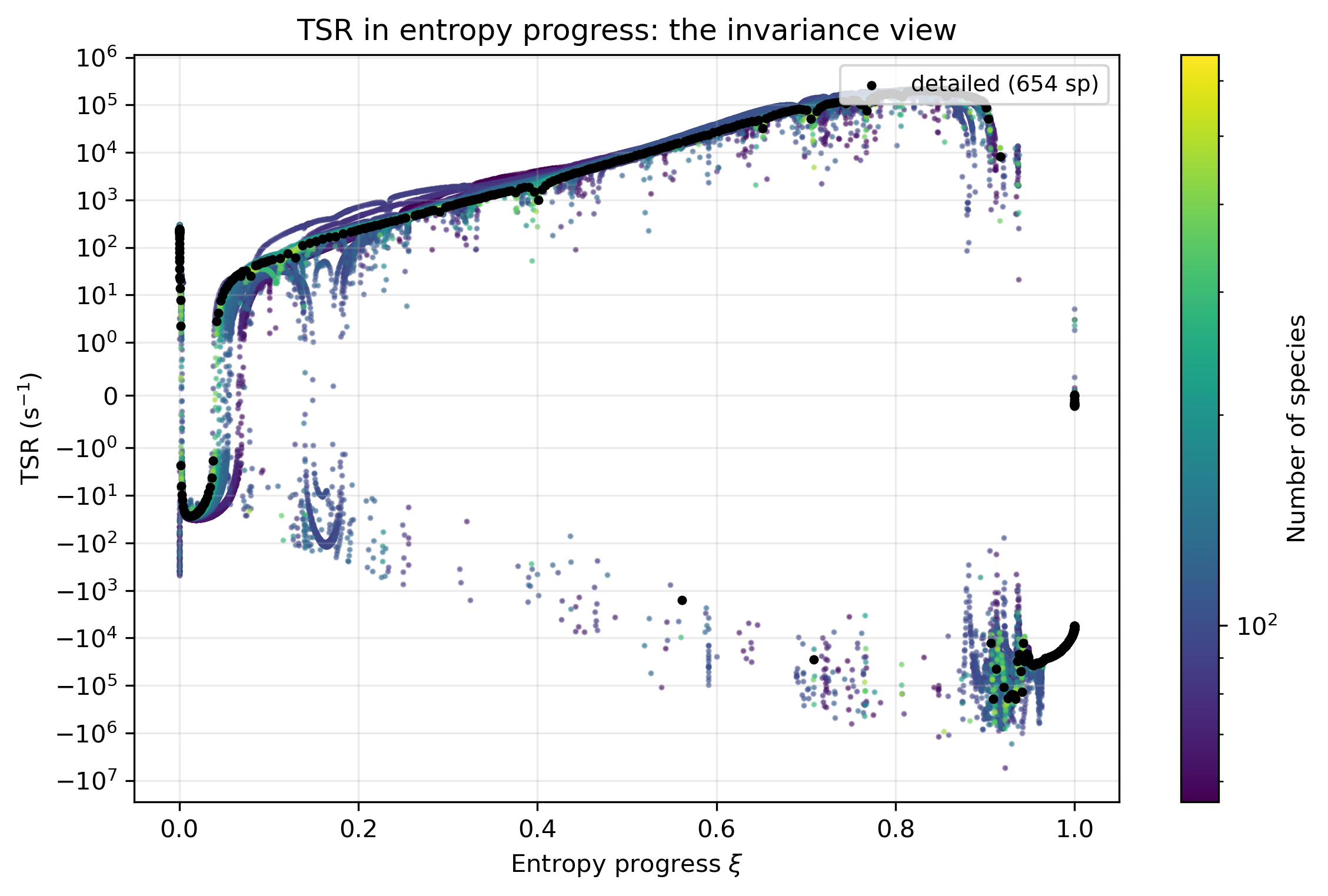}
\caption{TSR in normalised entropy progress: the ignition-region view.
All mechanisms ride the same backbone; the sparse cloud of
negative-TSR states beneath it is discussed in the text and in
Fig.~\ref{fig:tsr_pair}.}
\label{fig:tsr_xi}
\end{figure}

\paragraph{Invariance under skeletal reduction.} Since the TSR is a
property of the driving dynamics rather than of its representation, an
accurate skeletal mechanism should leave it essentially unchanged.
Figures~\ref{fig:tsr_time} and~\ref{fig:tsr_xi} test this ansatz on the
training datasets of the 33-mechanism family in two complementary
parametrisations. Logarithmic time (Fig.~\ref{fig:tsr_time}) expands the
\emph{induction} period: the TSR plateau at
$\omega_\tau \approx 2\times10^{2}$~s$^{-1}$ is tightly bundled across
all mechanisms over eight decades of time, and the common negative basin
just before ignition -- a genuine pre-ignition relaxation phase, present
in the detailed mechanism as well -- is clearly resolved. Normalised
entropy progress $\xi$ (Fig.~\ref{fig:tsr_xi};
Section~\ref{sec:accuracy}) instead expands the \emph{ignition} region,
where the entropy production concentrates: all mechanisms superpose on a
single TSR backbone -- induction plateau, monotone explosive ramp, and
the plunge to strong contraction past $\xi \approx 0.9$ -- and the
\emph{peak} stretching rate is invariant to $\sim$2\% across the entire
family ($\omega_\tau^{\max} = 2.00$--$2.05\times10^{5}$~s$^{-1}$ from 56
to 654 species), even for the smallest members whose ignition delay errs
by $\sim$60\%: the reduction may shift \emph{when} ignition happens, but
not \emph{how violently}.

\begin{figure}[htbp]
\centering
\includegraphics[width=0.8\textwidth]{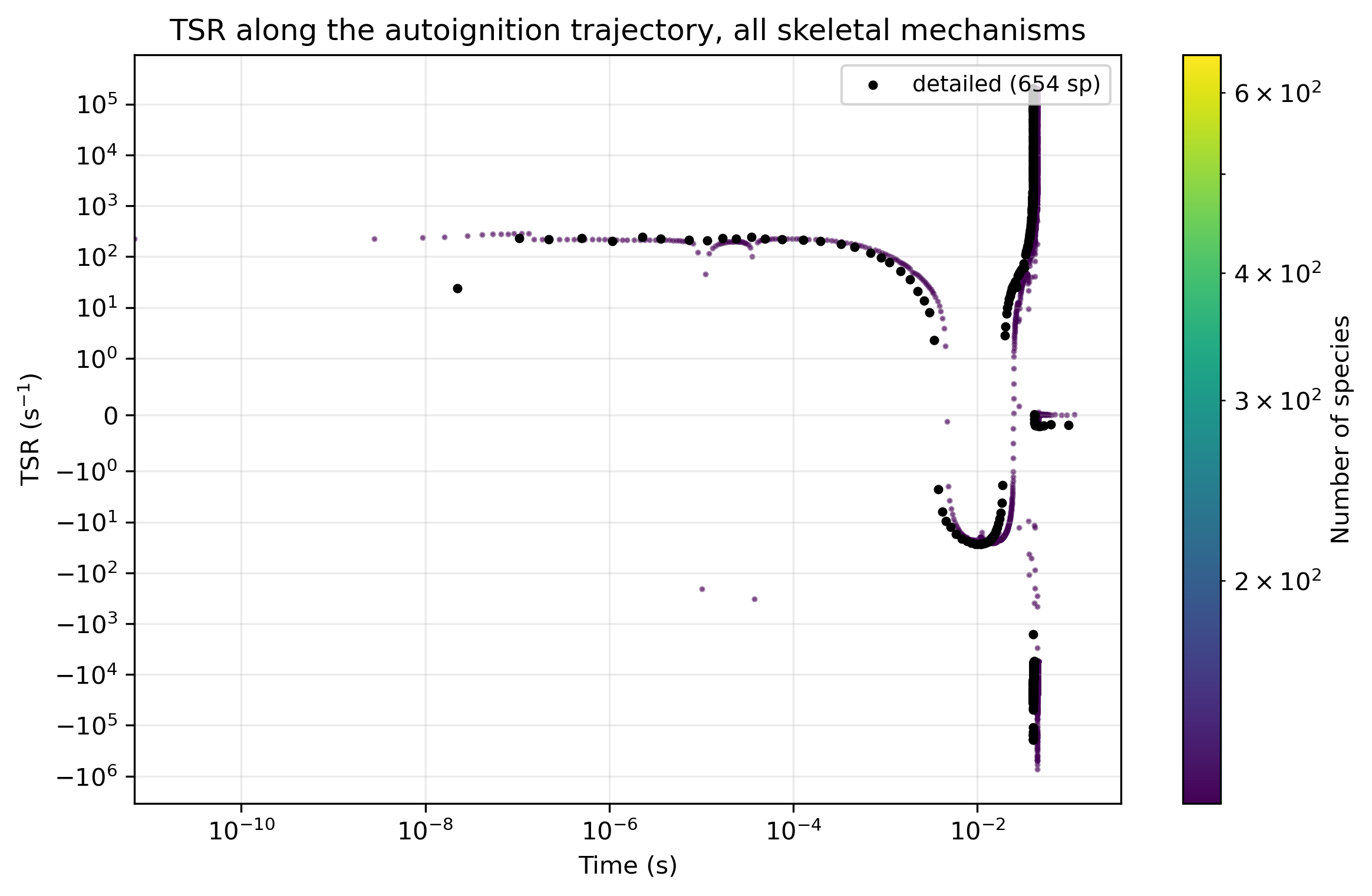}
\caption{TSR time histories of the 121-species skeletal mechanism and
the detailed 654-species mechanism. The two curves track each other
along the entire trajectory; the pre-ignition negative basin is shared,
and the reduced mechanism adds only two narrow oscillatory-episode dips
during the early transient.}
\label{fig:tsr_pair}
\end{figure}

\paragraph{Anatomy of the negative excursions.} The sparse cloud of
negative-TSR states beneath the backbone of Fig.~\ref{fig:tsr_xi}
deserves explanation, lest it be mistaken for numerical noise. Figure~\ref{fig:tsr_pair} isolates one skeletal member
(121 species) against the detailed mechanism: the two TSR histories
track each other along the entire trajectory -- including the
pre-ignition negative basin -- and the reduced mechanism adds only two
narrow negative dips of its own during the early transient. A per-state
dissection shows that such excursions are \emph{few, smooth and
multi-state} rather than isolated glitches (in the 121-species ramp, 229
of 1361 states have $\omega_\tau < 0$, of which only five are
single-state events and only one coincides with a change of the
exhausted-mode count $M$). Each episode corresponds to a \emph{weakly
damped complex-conjugate pair momentarily dominating the weights} (e.g.\
$\lambda \approx -1.2\ldots{-2.6} \pm 13\ldots16\,i$~s$^{-1}$ carrying
$\sim$45\% of the weight): by Eq.~\eqref{eq:tsr} the pair contributes
$\mathrm{sign}(\mathrm{Re}\,\lambda)\,|\lambda| = -|\lambda|$, so the
excursion's magnitude is set by the pair's \emph{frequency} rather than
its damping -- oscillatory interludes are rendered as prominent negative
excursions by construction of the index. The apparent population of the
negative cloud in the family overlay is then an aggregation effect: each
mechanism contributes a handful of such episodes at slightly shifted
$\xi$, and 33 superposed sparse sets read as a dense cloud. This is precisely why the
quantitative comparison below operates on the median-filtered backbone.

\begin{figure}[htbp]
\centering
\includegraphics[width=0.7\textwidth]{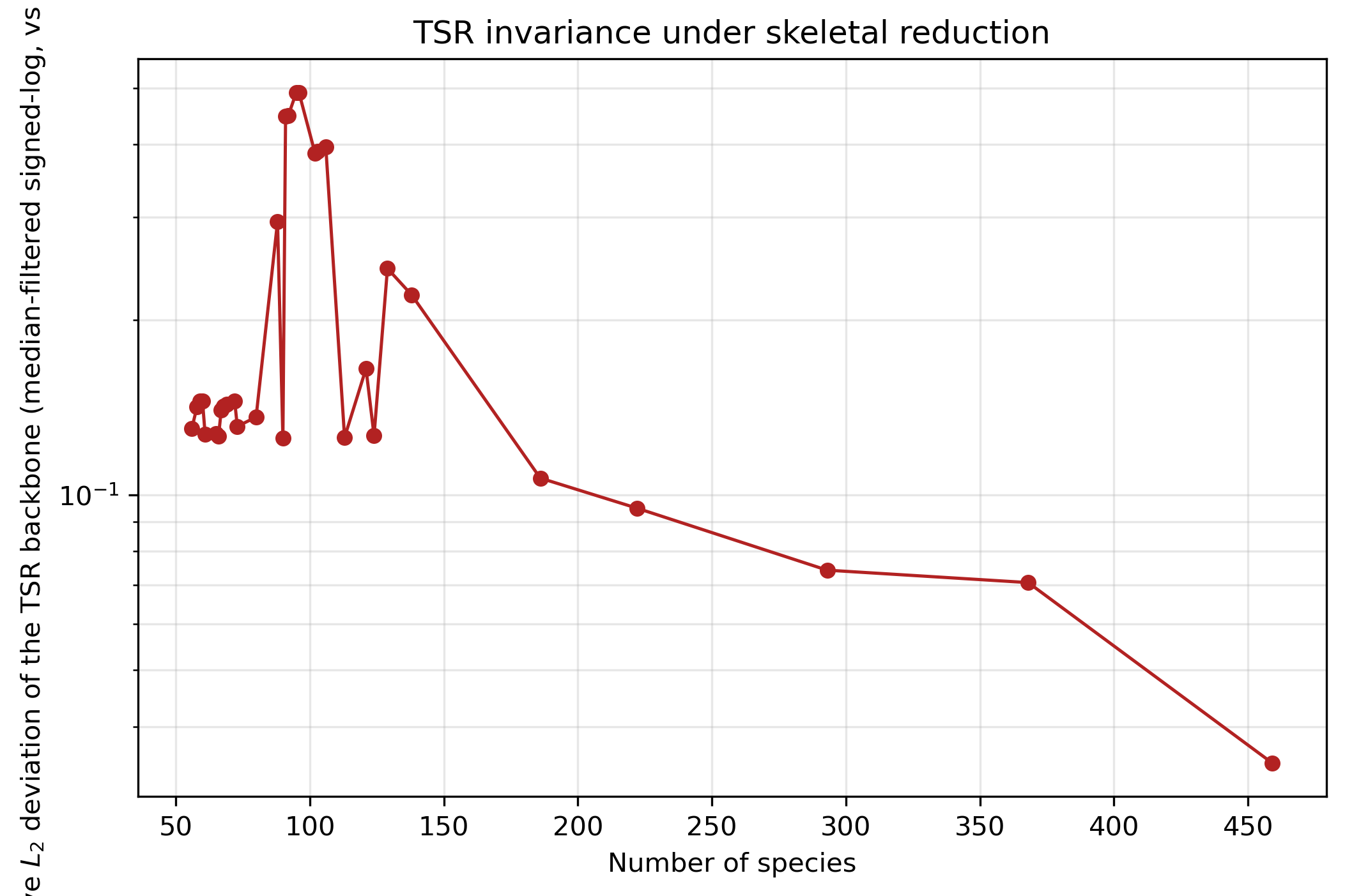}
\caption{Relative $L_2$ deviation of the TSR backbone from the detailed
mechanism, against mechanism size. The anomalous 88- and
91--106-species members coincide with those flagged by the
ignition-delay assessment (Section~\ref{sec:accuracy}).}
\label{fig:tsr_dev}
\end{figure}

\paragraph{A mechanism-quality diagnostic.} The deviation of each
mechanism's TSR backbone from the detailed one (relative $L_2$ of the
median-filtered signed-log profile over $\xi$) is shown in
Fig.~\ref{fig:tsr_dev}. It decays with mechanism size, to 0.035 at 459
species -- but not monotonically: the 91--106-species subfamily deviates
three- to four-fold more than its neighbours, and the 88-species member
is likewise elevated. These are \emph{precisely} the members flagged by
the ignition-delay assessment of Section~\ref{sec:accuracy} as a locally
anomalous stretch of the reduction sequence. Two independent metrics
thus identify the same irregularity -- and the TSR one is computable
from a single sampled trajectory per candidate mechanism, with no
reference integrations of the candidates. The TSR-profile deviation can
therefore serve as an inexpensive screening diagnostic when generating
skeletal families for hash-table (or any other) use, complementing the
global observables customarily used to certify a reduction. Looking
forward, the same quantity could be promoted from diagnostic to
\emph{design constraint}: a simplification algorithm that requires the
candidate mechanism to reproduce the baseline TSR evolution within a
prescribed deviation would certify dynamical fidelity -- the driving
time-scale structure -- rather than endpoint fidelity alone, and would
have rejected the anomalous members of the present family at generation
time.

The study is reproduced by \texttt{compare\_tsr.py}, distributed with
the code.

\end{document}